# Numerical study on wide gap Taylor Couette flow with flow transition

M. A. Razzak[1], B.C. Khoo[1] and K.B. Lua[2]

[1] Department of Mechanical Engineering, National University of Singapore, Singapore 119260, Singapore

[2] Department of Mechanical Engineering, National Chiao Tung University, Taiwan

**Abstract:** This study aims to investigate the possible sources of non-axisymmetric disturbances and their propagation mechanism in Taylor Couette flow (TCF) for wide gap problems using direct numerical simulation with a radius ratio of 0.5 and Reynolds number (Re) ranging from 60 to 650. Here, attention is focused on the viscous layer (VL) thickness in near-wall regions and its spatial distribution along the axial direction to gain an insight into the origin and propagation of non-axisymmetric disturbances. The results show that an axisymmetric Taylor-vortex flow occurs when Re is between 68 and 425. Above Re = 425, transition from axisymmetric to non-axisymmetric flow is observed up to Re = 575 before the emergence of wavy-vortex flow. From the variation of VL thickness with Re, the VL does not experience any significant changes in the flow separation region of the inner wall, as well as jet impingement region of both the inner and outer walls. However, a sudden increase in VL thickness in the flow separation region of the outer wall reveals possible source of non-axisymmetric disturbances in the flow separation region of the outer wall. These disturbances develop into the periodic secondary flow as the axisymmetric flow transforms into non-axisymmetric flow and this leads to the emergence of azimuthal wave. The periodic secondary flow contributes to sudden increase in the natural wavelength and rapid reduction in the strength of two counter-rotating Taylor vortices. This in turn leads to a substantial reduction of torque in the transition flow vis-a-vis axisymmetric Taylor-vortex flow.

## 1. Introduction

The study of the vortex structures in Taylor Couette flow (TCF) has gained a lot of attention from researchers partly due to its complexity and nonlinear behaviour which warrants careful investigation. Generally, TCF with a rotating inner cylinder and a stationary outer cylinder undergoes a series of flow transitions as Reynolds number (Re) increases. These flow transitions are classified into circular Couette flow, axisymmetric Taylor vortex flow (ATVF), periodic non-axisymmetric flow or wavy vortex flow (WVF), modulated wavy vortex flow (MWVF) and turbulent Taylor vortex flow (TTVF)

(Taylor, 1923; Synge, 1938; Chandrashekar, 1953; 1958; Donnely, 1958; Coles, 1965; Lim et. al 1998; Lim & Tan, 2004). The physical parameters which had been widely studied in TCF are radius ratio (inner radius/outer radius) and aspect ratio (height of cylinder/annular gap). These studies are categorized into the narrow gap and wide gap problems based on the radius ratio (Chandrashekar, 1953; 1958; Donnely, 1958). In the narrow gap problems, the annular gap (i.e., outer radius - inner radius) is much smaller than the mean radius and usually indicates that radius ratio > 0.5 (Chandrashekar, 1953; 1958; Donnely, 1958). For the wide gap problems, the radius ratio is $\leq 0.5$. From the physics point of view, the critical Re at which ATVF appears increases with increase in radius ratio in narrow gap problems while the opposite behaviour is observed for the wide gap problems (Sparrow, Munro & Jonsson, 1964).

Numerous studies have been conducted on the transition states of TCF for the narrow gap configuration. It was reported that the flow in the separation ( i.e., fluid moving away from wall) and jet impingement regions ( i.e., fluid moving towards wall) plays a significant role in the transition of ATVF into WVF and WVF into TTVF (Snyder & Lambert, 1965; Kataoka, Doi & Komai ,1977; Gorman & Swinney, 1979; Sobolik, Benabes & Cognet, 1995; Dumont et al., 2002; Akonur & Lueptow, 2003; Sobolik et. al., 2011; Kristiawan, Jirout & Sobolík, 2011; Gao, Kong & Vigil, 2016; Dessup et al., 2018).

Relatively fewer studies have been done on the wide gap problems. Although, Taylor (1923) had explained the flow stability and formation of toroidal vortices (Taylor vortices) for the narrow gap problems, the stability model developed by Taylor (1923) did not provide a good approximation for the wide gap problems (Synge, 1938; Chandrashekar, 1953; 1958 and Donnelly, 1958). This led Chandrashekar (1953; 1958) and Donnelly (1958) to conduct a linear stability analysis to investigate the onset of instability of ATVF for wide gap problems (i.e., radius ratio = 0.5). However, their studies were limited to measurement of torque and determination of the onset of instability for a small range of Re. These studies show that the critical Re at which ATVF appears is 68 and the WVF was considered to occur at a value 10 times greater than the critical Re. Subsequent works by Stuart (1957; 1960), Watson (1960) and Davey (1962) had identified that the appearance of the WVF is related to the nonlinear growth of non-axisymmetric disturbances ( i.e., disturbances that change along the azimuthal direction ), and thus their studies reckoned that it was necessary to capture the origin and propagation of non-axisymmetric disturbances. This could explain why Chandrashekar (1953; 1958) and Donnelly (1958) were unable to provide details about the WVF in their linear stability analysis (Stuart,1957; 1960; Watson, 1960; Davey, 1962). The experiments conducted by Donnelly & Simon (1960) and Donnelly & Fultz (1960) for the wide gap problem (radius ratio 0.5) and aspect ratio 5 showed good agreement with the study of Chandrashekar (1953; 1958) and Donnelly (1958). However, the use of relatively small aspect ratio of less than 8 in Donnelly & Simon (1960) and Donnelly & Fultz (1960) may have strongly influenced the appearance of WVF regime at a higher Re (Cole, 1974; Walden & Donnelly, 1979; Blennerhassett & Hall, 1979; Hall, 1980). In a later study, Fasel & Booz (1984) carried out direct numerical simulation (DNS) for the wide gap problem (radius ratio 0.5) using infinite length cylinders for Re ranging between 60 to 650 with the assumption of axisymmetric flow structure. The assumption of an axisymmetric flow structure in their simulation had contributed to the absence of WVF structure. Heise et al. (2008) and Abshagen et al. (2008) studied the nonlinear dynamics of ATVF and WVF in an attempt to understand several bifurcation processes exhibited in the TCF for wide gap problem. Manneville & Czarny (2009) studied the aspect ratio dependency of Taylor vortices for wide gap problem. Abshagen et al. (2012) also carried out experimental investigation on the multiplicity of states in TCF which appeared due to the axial localization of azimuthal travelling wave for wide gap problems. This led López et al. (2015) to conduct a numerical analysis in an attempt to better understand the multiplicity of states found in Abshagen et al. (2012). In Martiand et al. (2014), stability analysis had been used in the ATVF and WVF flow for radius ratio 0.55 and Re ranging between 69 to 210. This study had reported that the cause of the transition between ATVF into WVF is not the same in the narrow gap and wide gap problem. In the wide gap problem, the subharmonic modes of instability or jet mode were the principal cause of the appearance of the WVF. The weakly nonlinear analysis was implemented in Martiand et al. (2017) to study the effect of additional external radial and axial flow along with the rotation of the inner wall for a wide range of radius ratio. Though Martiand et al. (2014; 2017) had provided extensive details about the stability analysis of TCF for Re ranging from 69 to 210 in wide gap problems, clear understanding of the physical mechanism of the transition of ATVF into WVF and how the flow behaves at higher Re in wide gap problem yet to explore. In addition, the source of non-axisymmetric disturbances, their propagation mechanism and influence of these disturbances into the natural wavelength and normalized torque yet to be understood clearly in TCF with a wide gap.

Although these earlier studies had provided much insights on the axisymmetric flow and the multiplicity of states observed in non-axisymmetric flow structure, most of them were focused on the implementation of nonlinear stability model for the wide gap problems. This motivates us to conduct a detailed study of the physical behaviour exhibited in the transition from ATVF into WVF flow structure of wide gap problems. Earlier studies of small gap problems had indicated that the flow behaviour in the flow separation and jet impingement regions are critical to the origin and propagation of non-axisymmetric disturbances as they eventually triggers the onset of the periodic non-axisymmetric flow structure (i.e., WVF) ; see Sobolik, Benabes & Cognet (1995), Dumont et al. (2002) , Akonur & Lueptow (2003); Sobolik et. al. (2011), Kristiawan, Jirout & Sobolík (2011) and Dessup et al. (2018). These observation prompted us to investigate in details how the source of non-axisymmetric disturbances and their physical propagation mechanism vary between small gap and wide gap problem during the transition of ATVF into WVF and WVF at higher Re ( i.e., Re = 60 to 650) with the aid of spatial behaviour of near-wall regions (i.e., the region where azimuthal velocity varies linearly with the distance from wall) along the axial direction . The study of the near-wall region in open boundary flow and internal flow is a well-established concept (Pope, 2000). In particular, in the open boundary or internal flow, the dynamics of vortices are restricted by the wall on one side while the other side is open to the fluid flow exposed to flow separation or jet impingement, that occurs only in the near-wall region of the junction between two counter-rotating vortices (i.e., streamwise vortices, longitudinal vortices, Dean vortices, Gortler vortices as documented in Wu, Ma, & Zhou, 2006). However, in the TCF problems, Taylor vortices are confined between two walls that lead to the appearance of both jet impingement and flow separation at the near-wall regions of same junction ( i.e., inflow or outflow regions) of two counter-rotating vortices (Dumont et al., 2002 ; Akonur & Lueptow, 2003; Sobolik et. al., 2011; Kristiawan, Jirout & Sobolík, 2011). A study of the near-wall regions in the TCF may be the key for a fuller and better understanding of the underlying flow physics. The only few studies carried out on the near-wall regions in TCF are for TTVF of small gap problems (Huisman, et al., 2013; Rodolfo et al., 2016 and Singh et al., 2016). Broadly, the study of the near-wall regions in ATVF, WVF and MWVF in both the wide and small gaps problems are yet to be fully explored. In particular, this study focuses on the behaviour of viscous layer (VL) thickness in the near-wall regions of ATVF and WVF in wide gap problems (radius ratio 0.5) and concentrates on the flow separation and jet impingement regions.

In this study, direct numerical simulation (DNS) has been employed to solve three dimensional TCF flow with a radius ratio of 0.5. For the purpose of validating the numerical tools used in this present study, a single wavelength fluid column with periodic boundary condition in the axial direction is conducted and compared against the results of Fasel & Booz (1984). Following this, with the advancement of computational resources, the axial height of cylinder is extended to four wavelength of fluid column (8 times the annular gap size) with periodic boundary condition at the both ends of cylinder to investigate how the flow structure behaves with the change in axial height of fluid column from single wavelength to four wavelengths. The use of four wavelengths fluid column implies an aspect ratio of 8. As reported in Cole (1974), the average NNW of Taylor vortex in wavy vortex flow was found as 1.98 for aspect ratio 8 and 2 for aspect ratio 44 with end effect. Which indicates that 1% deviation in NNW between aspect ratio 8 and 44. Blennerhassett & Hall (1979 ) had reported that the modes of instability become independent of aspect ratio if it was $\geq 6$ with the end effect. In the present study, the aspect ratio of 8 has been used with a periodic boundary condition at both ends of the cylinder, and thus satisfies the assumption of the infinite length of the cylinder ( i.e., insignificant end effect).

To aid the analysis of the dynamics of near-wall regions in the flow separation and jet impingement on both inner and outer walls, we define a viscous layer (VL) and examine its spatial behaviour along the axial direction as Re changes. In doing so, the analysis provides an indication on the transition from axisymmetric to non-axisymmetric flow structure, and in turn, helps to elucidate the origin of periodic non-axisymmetric disturbances (i.e., WVF) and their propagation. Additionally, the study of VL with Re enables us to examine the influence of variation of natural wavelength on the flow dynamics within the near-wall region. Finally, the correlation between the distribution of VL along the axial direction and local wall shear stress at the jet impingement and flow separation regions are also studied. The behaviour of VL thickness in the transition from ATVF to WVF flow structure and its dependency on the natural wavelength may provide useful clue to the possible mechanism that leads to drag reduction.

## 2. Flow configuration

The schematic diagram of the TCF set up is illustrated in Figure 1 and unless otherwise stated, cylindrical coordinates (r, θ, z) are adopted with the z-axis coinciding with the axis of symmetry of the cylinders. The radii of the inner and outer cylinders are $R_i$ and $R_o$, respectively. The annular gap is expressed as $d = R_o - R_i$ and the axial height of the cylinders or fluid column is represented as $H$. The non-dimensional radius ratio is defined by $\frac{R_i}{R_o}$. Here, the inner cylinder is in rotation with a constant angular velocity Ω while the outer one is kept stationary. The fluid in-between the two concentric cylinders is considered to be viscous and incompressible. The Reynolds number (Re) is expressed with the following equation,

$$\text{Re} = \frac{\Omega R_i d}{\upsilon} \qquad (1)$$

where $\upsilon$ is kinematic viscosity of fluid.

## 3. Numerical Method and Validation test

Most of the previous numerical studies on TCF were conducted using a single wavelength ( two gap widths or aspect ratio 2) fluid column as the axial height with periodic boundary condition at the both ends of cylinder (Marcus, 1984; Fazel & Booz, 1984; Liao, Jane & Young, 1999; Bilson & Bremhorst, 2006; Dong, 2007; Abshagen et al., 2008; Pirrò & Quadrio, 2008; Heise et al., 2008; Jeng & Zhu, 2010; Kristiawan, Jirout & Sobolík, 2011; Sobolik et al., 2011; López et al., 2015; Ng, Jaiman & Lim, 2018; Dessup et al., 2018). In this study, DNS has been carried out for two different heights of fluid column, namely single wavelength fluid column (Height = 2 times the annular gap size) and four wavelengths fluid column (Height = 8 times the annular gap size). In both cases, the radius ratio is 0.5, Re ranges between 60 to 650 and periodic boundary condition at the both ends of cylinder is used as mentioned in the above studies. As mentioned in the previous section, a single wavelength fluid column is carried out to validate our numerical method against the results of Fasel & Booz (1984), whereas the four wavelengths fluid column is used to investigate the variation of natural wavelength and spatial distribution of VL along the axial direction with respect to Re.

A structured mesh (Hexahedral) is generated using Fluent Mesher, and the growth rate of the prism layer from the wall is set as 5. This provides a non-uniform distribution of grid size along the radial direction with the region near the wall having a finer grid size and the other regions having a relatively larger grid size.

DNS is conducted by solving the three dimensional Navier-Stokes equation by finite volume method using icoFoam solver in OpenFOAM 5. As per incompressible flow, the pressure based solver has been used. Interpolation scheme without non-orthogonal correction was used to compute the cell face pressure. The second order discretization scheme without non-orthogonal correction was applied for the convective, diffusive fluxes and velocity derivatives. A second order implicit method (Backward) was used for the transient formulation. The Pressure Implicit with Splitting of Operators (PISO) was used to obtain the pressure correction. Generalized geometric-algebraic multi-grid (GAMG) solver along with Smoother Gauss-Seidel was imposed to solve for the pressure while the Preconditioned (bi-) conjugate gradient (PBiCG) linear solver with Diagonal incomplete-LU (asymmetric DILU) was used to solve for the velocity. From the stability criterion, Courant–Friedrichs–Lewy (CFL) < 0.5 was used for the whole range of Re. As reported in the previous study (Dessup et al., 2018), the viscous time scale required for TCF flow to stabilize is $\frac{d^2}{v}$ (i.e., d is annulus gap and $v$ is kinematic viscosity). In the present study, the radius of the inner cylinder is 0.0015m and the radius of the outer cylinder is 0.003m. This provides a gap of 0.0015m between two cylinders. The kinematic viscosity used here is $1.0048 \times 10^{-6} m^2 s^{-1}$ which translates the viscous time scale for our problem to be 2.239252 sec. The simulation time (i.e., flow time) used in this study is 5 sec which is 2.239 times greater than the required viscous time scale. This indicates that our flow structure is stabilized within the time used in the simulation.

For expediency, the whole range of Re has been divided into four regions to carry out the grid independence test (see Table 1). For each region, the maximum Re is used as the reference for the grid independence test (see Figure 2) and the number of grid points along the radial, azimuthal and axial direction is summarized in Table.1. The normalize torque presented in Figure 2, is obtained with the following equation,

$$Normalized\ Torque,\ \bar{\tau} = \frac{T}{v\rho\Omega R_i^2 H} \quad (2)$$

where $T$ is the torque and $\rho$ is the density of fluid.

The torque deviation between the inner and outer cylinder was observed to be less than 0.4% which indicates the finer scale of mesh along the radial direction.

Table 1: Number of grid points along radial, azimuthal and axial directions

| Re | Radial direction | Azimuthal direction | Axial direction (Each wavelength) |
|---|---|---|---|
| 60≤Re≤175 | 70 | 250 | 125 |
| 175≤Re≤275 | 85 | 300 | 150 |
| 275≤Re≤475 | 100 | 350 | 175 |
| 475≤Re≤650 | 115 | 400 | 200 |

The DNS result presented in Figure 3 for a single wavelength fluid column shows excellent agreement with the DNS study of Fasel & Booz (1984) up to Re = 425 with a maximum deviation of 0.12%. The critical Re obtained in this study is 68 which is essentially the same as Fasel & Booz (1984) . However, as Re increases beyond 425, the discrepancy between the present DNS of a single wavelength fluid column and Fasel & Booz (1984) increases (see Figure 3). In Fasel & Booz (1984), the flow structure is assumed to be axisymmetric and they considered the Re at which non-axisymmetric flow structure appears is nearly 10 times greater than critical Re (i.e., Re = 680). Our DNS study does not assume any axisymmetric flow structure. The streamlines associated with the flow structure obtained in the r-Z plane at two azimuthal locations ( i.e., θ = 0 and 180 degrees) are illustrated in Figure 4. The flow structure obtained at θ = 0 and 180 degree azimuthal locations are identical up to Re 425, beyond which a shift of inflow region (i.e., junction between two Taylor vortices where fluid moves from outer wall towards the inner wall) along the axial direction between these two locations is observed (see Figures 4 d-f). This suggests that the flow structure is axisymmetric for Re ranging from 60 to 425 (see Figures 4 a-c) and beyond which increasingly non-axisymmetric flow structure is observed (see Figures. 4 d-f). This probably explains the increase in the deviation between the present DNS for single wavelength fluid column and that of Fazel & Booz (1984) as Re increases beyond 425. Thus, the deviation observed in Figure 3 provides an indication of the appearance of the non-axisymmetric flow structure (see Figures 4 d-f). Separately, a DNS calculation for the single wavelength fluid column with axisymmetric assumption has been especially carried out for Re 650 and it is found that our DNS of the single wavelength fluid column result is indeed very close to that of Fasel & Booz (1984) with a discrepancy of 0.39 %. Therefore, the numerical tools used in the present DNS study has exhibited good validation against Fasel & Booz (1984).

The DNS results obtained for four wavelength fluid column is discussed in the following section.

## 4. On four wavelength fluid column configurations

### 4.1 Variation of flow regimes

The azimuthal wall shear stress of the outer wall has been non-dimensionalized as follows,

$$\text{Normalized azimuthal shear stress at the outer wall, } \bar{\tau}_{r\theta}^{o} = \frac{\tau_{r\theta}^{o} d}{v \Omega \rho R_i} \qquad (3)$$

where $\tau_{r\theta}^{o}$ is azimuthal shear stress at outer wall.

As shown in Figure 5, the distribution of $\bar{\tau}_{r\theta}^{o}$ at the outer wall along the azimuthal direction (0 to 360 degrees) obtained for the four wavelengths fluid column exhibits uniform distribution up to Re = 425 in the inflow region (see Figure 5a) and outflow region (Figure 5b) (i.e., junction of two Taylor vortex where the radial flow is moving from inner wall towards outer wall). This suggests that the flow structure is independent of the azimuthal location. As Re is increased beyond 425, the distribution of $\bar{\tau}_{r\theta}^{o}$ reveals a wavy like behaviour in the inflow (see Figure 5a) and outflow regions (see Figure 5b). This indicates that the $\bar{\tau}_{r\theta}^{o}$ changes along the azimuthal direction, thus indicating the emergence of non-axisymmetric flow structure when Re exceeds 425. The streamline patterns in the r-Z planes at two azimuthal locations ( i.e., θ = 0 and 180 degrees) as presented in Figure 6 illustrate that the flow structure at θ = 0 and θ =180 are similar up to Re = 425 (see Figures 6 a-d); beyond which clear deviation is observed (see Figures 6 e-h). This also confirms the appearance of non-axisymmetric flow structure at Re > 425. It was mentioned in the earlier section that the circular Couette flow transforms into axisymmetric Taylor vortex flow (i.e., ATVF) at Re = 68. Thus, it is obvious that ATVF flow regime ranges between Re = 68 to Re = 425. This finding partly contradicts the Re range of axisymmetric flow structure obtained by Chandrashekar (1958), Donnelly (1958), Donnelly & Simon (1960), Donnelly & Fultz (1960) and Fasel & Booz (1984). The inconsistency could be attributed to the inability of their linear stability theory in capturing the growth and propagation of nonlinear disturbances (Stuart, 1960; Watson, 1960; Davey,1962). Although Donnelly & Simon (1960) and Donnelly & Fultz (1960) reported good agreement between experiment and the linear stability theory of Chandrashekar (1953; 1958) and Donnelly (1958), the appearance of non-axisymmetric flow structure in their study may have been delayed as they have used a smaller aspect ratio of less than 8 (Cole, 1974; Walden & Donnelly, 1979; Blennerhassett & Hall, 1979; Hall, 1980). Using nonlinear dynamics analysis, Abshagen et al. (2008) noted the appearance of non-axisymmetric flow structure at Re ≥ 400 for radius ratio = 0.5. Hence, the appearance of non-axisymmetric flow structure (see Figure 5) observed in this study could be justified by the findings of Stuart (1960), Watson (1960), Davey (1962), Coles (1974) and Abshagen et al. (2008).

The results in Figure 5 also show that at the beginning of non-axisymmetric flow structure, the distribution of normalized wall shear stress indicates the azimuthal wavenumbers (i.e. the number of wavelengths along azimuthal direction) is 2 (see Figure 5a and Figure 5b) for Re = 475, 525 and 575. Martiand et al. (2014) had observed 2 azimuthal wavenumbers at the beginning of non-axisymmetric flow structure for radius ratio 0.55 and 0.65. Thus, the azimuthal wavenumber observed in the present study at the beginning of non-axisymmetric flow structure is in alignment with Martiand et al. (2014). For Re = 600 and 650, the azimuthal wavenumber is found to be 1 in the present study (see Figure 5a and Figure 5b) which is in good agreement with Abshagen et al. (2008; 2012).

The normalized area average axial velocity ( $\bar{V}_z$ ) in the r-Z plane at θ location has been calculated using the following equation,

$$\bar{V}_z = \frac{\int_{R_i}^{R_o} \int_0^H V_z(z) dz dr}{(R_o - R_i)\Omega R_i H} \quad . \tag{4}$$

The average axial velocity at θ = 0, 90, 180 and 270 in the ATVF flow regime is found to be nearly zero (see Figure 7). At the beginning of the emergence of non-axisymmetric flow structure (i.e., Re = 425), a rather sudden increase in the magnitude of $\bar{V}_z$ is observed (see Figure 7). As shown in Figures 6 a-d, in ATVF flow regime (i.e., Re ≤ 425), there is no axial flow in the inflow and outflow region of Taylor vortices. However, as Re increases to 475, the presence of axial flow in the inflow and outflow regions of Taylor vortices (see Figures 6 e-f) is observed which is likely related to the sudden increase in the magnitude of $\bar{V}_z$ and it changes periodically over time. The observed non-zero $\bar{V}_z$ between two adjacent Taylor vortices is defined here as periodic secondary flow (see Figures 6 e-f). Although this flow appears in both inflow and outflow regions, it exhibits a more prominent behaviour in the inflow region. $\bar{V}_z$ which is due to the presence of periodic secondary flow, increases with Re and reaches its maximum value at Re = 525 for θ = 0 and180 and Re = 475 for θ = 90 and 270. Beyond Re = 475, a large drop is observed until Re = 575 to be followed by a more gradual drop until Re = 650 (see Figure 7). This suggests strong periodic secondary flow up to Re ≈ 575 before the strength weaken. Based on the behaviour of azimuthal wavenumber (see Figure 5) and periodic secondary flow (see Figure 6), the non-axisymmetric flow structure can be divided into two regimes, namely, transitional flow regime of ATVF into WVF and WVF flow regime. The former ranges between Re > 425 to Re < 575 and the latter is observed at Re ≥ 575.

From azimuthal vorticity ($\omega_\theta$) contour at the middle of annular gap illustrated in Figure 8, it is found that inflow and outflow regions of Taylor vortices exhibit wavy behaviour along the azimuthal direction at Re ≥ 425. Despite both the flow regions exhibiting wavy behaviour along the azimuthal direction, the inflow region has a far greater wave amplitude than that of outflow region (see Figures 8 b-f); this is due to comparatively stronger periodic secondary flow observed in the inflow region (see Figures 6e-f). Moreover, the azimuthal wave amplitude in the transition of ATVF into WVF is greater than that of WVF (see Figures 8 b-f).

From the discussions above, it is noted that the non-axisymmetric flow structure appears at the same Re for both the single wavelength and four wavelength fluid column (see Figures 4 and 6). Despite the similarity in Re, there is a distinct difference in the behaviour of their non-axisymmetric flow structure. As discussed in the earlier section, Fasel & Booz (1984) ) and Abshagen et al. (2008) had used the equivalent of a single wavelength as the fluid height, which essentially fixed the wavelength of Taylor vortices for all Re. This undoubtedly limits the natural behaviour and development of non-axisymmetric flow structure for the single wavelength fluid column. This could have contributed to the deviation in the behaviour of their non-axisymmetric flow structure compared to that of four wavelengths fluid column. In the four wavelengths fluid column, the natural wavelength of Taylor vortices is dependent on Re for both axisymmetric and non-axisymmetric flow structure. The investigation of four wavelengths fluid column has allowed us to identify two distinct flow regimes (i.e., transition of ATVF into WVF and WVF) in the non-axisymmetric flow. There are interesting details on the physical behaviour of flow structure, azimuthal wavenumber, azimuthal wave and periodic secondary flow in these flow regimes. These will be discussed in the following sections.

## 4.2 The variation of the strength of counter-rotating Taylor vortices

The normalized area average clockwise azimuthal vorticity ($\bar{\omega}_\theta^+$) and counter-clockwise azimuthal vorticity azimuthal vorticity ($\bar{\omega}_\theta^-$) in the r-Z plane at θ location are computed via,

$$\bar{\omega}_\theta^+ = \frac{\int_{R_i}^{R_o} \int_{H_{min}^+}^{H_{max}^+} \omega_\theta^+(z) dz dr}{(R_o - R_i)\Omega \int_{H_{min}^+}^{H_{max}^+} dz} \tag{5}$$

$$\bar{\omega}_\theta^- = \frac{\int_{R_i}^{R_o} \int_{H_{min}^-}^{H_{max}^-} \omega_\theta^-(z) dz dr}{(R_o - R_i)\Omega \int_{H_{min}^-}^{H_{max}^-} dz} \tag{6}$$

The variation of $\bar{\omega}_\theta^+$ and $\bar{\omega}_\theta^-$ of Taylor vortices with Re at the r-Z planes at θ = 0, 90, 180 and 270 are illustrated in Figure 9. The $\bar{\omega}_\theta^+$ and $\bar{\omega}_\theta^-$ of Taylor vortices are identical in ATVF flow regime (i.e., Re ≤ 425) (see Figures 9 a-d) which reveals two counter-rotating Taylor vortices are in equilibrium condition in ATVF flow regime. In addition, $\bar{\omega}_\theta^+$ and $\bar{\omega}_\theta^-$ of Taylor vortices in this ATVF flow regime increase linearly with Re (see Figures 9 a-d). On the other hand, $\bar{\omega}_\theta^+$ and $\bar{\omega}_\theta^-$ of Taylor vortices in the transition regime from ATVF into WVF (i.e., Re > 425) at first decreases with the increase in Re until a local minimum is reached at Re = 525 before they increase till the end of the transition flow regime (i.e., Re = 575) (see Figure 9). The slight difference between the $\bar{\omega}_\theta^+$ and $\bar{\omega}_\theta^-$ in this transition flow regime suggests slight imbalance of the two counter-rotating Taylor vortices (see Figures9 a-d). As described in the previous section, strong periodic secondary flow is observed between two counter-rotating Taylor vortices (see Figure 7) during the transition of ATVF into WVF. This secondary flow carries momentum flux from adjacent Taylor vortices along the axial direction which subsequently leads to the reduction in the strength of two counter-rotating Taylor vortices. Moreover, the emergence of slightly non-equilibrium two counter-rotating Taylor vortices in the transition is also related to the appearance of strong periodic secondary flow. In the WVF flow regime, increase in Re causes $\bar{\omega}_\theta^+$ of Taylor vortices (see Figure 9) to increase slightly in θ = 0 and 180 (see Figures 9 a and c) and decrease slightly in θ = 90 and 270 (see Figures 9 b and d). In contrast, higher Re causes $\bar{\omega}_\theta^-$ to decrease slightly in θ=0 and 180, and a steady increase in θ = 90 and 270 (see Figures 9 b and d). The variation $\bar{\omega}_\theta^+$ and $\bar{\omega}_\theta^-$ in θ = 0, 90, 180 and 270 in the WVF flow regime could be atributed to the azimuthal wave.

## 4.3 Variation of the natural wavelength of Taylor vortices

The behaviour of the natural wavelength of Taylor vortices with Re in the ATVF, the transition of ATVF into WVF and WVF is discussed in this section. The normalized natural wavelength of Taylor vortices are computed using the following equation (Lim & Tan, 2004),

$$\text{Normalized natural wavelength (NNW)} = \frac{2H}{Nd} \tag{7}$$

where N is the number of Taylor vortices.

The variation of the total number of Taylor vortices and its corresponding normalized natural wavelength (NNW) with Re are presented in Figure 10 and Figure 11, respectively. For the case of single wavelength fluid column, two cells of Taylor vortices (see Figure 4) are observed which gives NNW of 2.00 for Re between 68 and 650. For the case of four wavelengths fluid column and in the ATVF flow regime, 10 cells of Taylor vortices are observed at Re = 125, 175, 225 and 325 (see Figure 10) with NNW ≈1.60 (see Figure 11). This is nearly 20% smaller than the NNW of single wavelength fluid column (see Figure 12). Also, in the same ATVF flow regime, only 8 cells of Taylor vortex are found at Re = 275, 375 and 425 (see Figure 10) with the corresponding NNW ≈2.00 (see Figure 11). This is identical to that observed in the single wavelength fluid column (see Figure 12). In the transition of ATVF into WVF, the number of Taylor vortices reduces to 6 (see Figure 10) which translates into NNW≈2.67 (see Figure 11). The NNW in this transition flow regime for four wavelengths fluid column is nearly 33% greater than that of single wavelength fluid column (see Figure 12). This could be due to the presence of periodic secondary flow (see Figure 7) as highlighted in the earlier section (see Figures 6 e-f). In WVF flow regime, there is a total of 8 Taylor vortex cells (see Figure 10) with the corresponding NNW = 2.00 (see Figure 11); i.e. the same as that of single wavelength fluid column (see Figure 12). In the WVF flow regime, (NNW) observed in our study is 2 which is exactly the same as the experiment of Abshagen et al. (2012) on wide gap problem (i.e., radius ratio 0.5).

### 4.4 The variation of normalized torque

In Figure 13, the variation of $\bar{\tau}$ with Re for four wavelengths fluid column is compared against the result of Fazel & Booz (1984) for single wavelength fluid column. It can be seen from the figure that $\bar{\tau}$ for the four wavelengths fluid column exhibits good agreement with that of single wavelength fluid column (see Figure 13) in ATVF (i.e., $68 \leq Re \leq 425$) and WVF (i.e., $Re \geq 575$) flow regimes with a maximum of 3% difference (see Figure 14). In the transition of ATVF into WVF (i.e., 425 < Re < 575) flow regime, $\bar{\tau}$ is nearly 11% smaller than that of single wavelength fluid column (see Figure 14). Mangiavacchi, Jung & Akhavan (1992), Laadhari, Skandaji & Morel (1994), Baron and Quadrio (1996), Choi & Graham (1998) and Choi and Clayton (2001) had reported the appearance of periodic secondary flow along the spanwise direction as the primary cause of the reduction in wall shear stress under the application of spanwise oscillation. Similar to the above cited studies, the present investigation also shows a strong periodic secondary flow along the axial direction in the transition of ATVF into WVF flow regime (see Figure 7). This secondary flow could have caused Taylor vortices to weaken (see Figure 9), and in turn led to the reduction in $\bar{\tau}$ in the transition between ATVF into WVF flow regime.

### 5. Variation of the spatial behaviour of VL in the different flow regimes

The above findings show several salient flow features of wide gap problems in TCF with four wavelengths fluid column, namely the appearance of strong periodic secondary flow, comparatively weaker Taylor vortices, sudden increase in the NNW and occurrence of the azimuthal wave during the transition of ATVF into WVF flow regime. The influence of these flow features on the reduction of normalized torque in the transition flow regime is another significant observation in this investigation. Past studies of narrow gap problems show that near-wall regions of inflow and outflow regions play a crucial role in the eventual outcomes of TCF (Akonur & Lueptow, 2003; Kristiawan, Jirout & Sobolík, 2011; Sobolik et al., 2011). Along the same line of investigation as small gap problems, we take a closer look at the dynamics of the near-wall region of both the inner and the outer walls with the aids of the analysis of the viscous layer (VL) thickness. In particular, we would like to examine the spatial behaviour or distribution of the VL thickness along the axial direction in the r-Z plane. Though VL is an established concept in open boundary and channel flow (Pope, 2000), it has not been widely applied for TCF. As described in the earlier section, Huisman et al. (2013), Rodolfo et al., (2016) and Singh et al. (2016) had studied the relationship between non-dimensional wall velocity, wall distance and VL at a fixed axial location for small gap problems in Turbulent Taylor vortex flow. However, the investigation of VL thickness in Taylor vortex flow (i.e., ATVF), wavy vortex flow (i.e., WVF), modulated wavy vortex flow (i.e., MWVF) and the spatial behaviour of VL thickness along the axial direction in these flow regimes have yet to be fully explored. This section will detail such an investigation for the wide gap problems.

In this analysis, we have computed non-dimensional wall velocity $(u^+)$ and wall distance $(y^+)$ in the r-Z plane at $\Theta = 0$ and $180$ along the axial direction of four wavelength fluid column for ATVF, transition between ATVF into WVF and WVF flow regimes with the aid of the following equations for the inner and outer walls (Huisman, et al. ,2013; Rodolfo et al., 2016 and Singh et al., 2016) :

Non-dimensional velocity for the inner wall, $u_i^+(z) = \frac{V_\theta(R_i)}{u_{\tau_i}(z)} - \frac{V_\theta(r,\ z)}{u_{\tau_i}(z)}$ (8)

Non-dimensional wall distance from the inner wall, $y_i^+(z) = \frac{(r-R_i)u_{\tau_i}(z)}{\nu}$ (9)

Non-dimensional velocity for outer wall, $u_o^+(z) = \frac{V_\theta(r,\ z)}{u_{\tau_o}(z)}$ (10)

Non-dimensional wall distance from the outer wall, $y_o^+(z) = \frac{(R_o-r)u_{\tau_o}(z)}{\nu}$ (11)

Friction velocity at the inner wall, $u_{\tau_i}(z) = \sqrt{\frac{\tau_{r\theta(z)}^i}{\rho}}$ (12)

Friction velocity at the outer wall, $u_{\tau_o}(z) = \sqrt{\frac{\tau_{r\theta}^0(z)}{\rho}}$ . (13)

The peak value of $y^+(z)$ within the linear region where $u^+(z) \approx y^+(z)$, provides the VL thickness along the axial direction of the r-Z plane at a fixed $\Theta$ location. The spatial distribution of normalized

VL thickness along the axial direction in the inner and outer walls has been computed from the peak value of $y^+(z)$ and equations are shown below.

The normalized thickness of VL at the inner wall, $\overline{VL}_i(z) = \dfrac{y^+_{peak_i}(z)\nu}{u_{\tau_i}(z)d}$ (14)

The normalized thickness of VL at the outer wall, $\overline{VL}_o(z) = \dfrac{y^+_{peak_o}(z)\nu}{u_{\tau_o}(z)d}$ (15)

The normalized azimuthal shear stress at the inner wall, $\bar{\tau}^i_{r\theta} = \dfrac{\tau^i_{r\theta}(z)d}{\nu\Omega\rho R_i}$ (16)

The normalized azimuthal wall shear stress at the outer wall, $\bar{\tau}^o_{r\theta} = \dfrac{\tau^o_{r\theta}(z)d}{\nu\Omega\rho R_i}$. (17)

The normalized axial height, $\overline{H} = \dfrac{Z}{H}$ (18)

Here, $V_\theta(r,z)$ is azimuthal velocity, $\tau^i_{r\theta}$ is azimuthal wall shear stress at inner wall and $\tau^o_{r\theta}$ is azimuthal wall shear stress at outer wall.

In laminar TCF flow, toroidal vortex structure occupies the annulus gap along the axial direction. The azimuthal velocity profile (Figure 15a) at different axial locations at the r-Z plane of Θ=0 location provides a clear idea about the influence of vortex structure on the azimuthal velocity distribution along the radial direction. As shown in Figure 15a, azimuthal velocity profile along the radial direction exhibits approximately two linear regions, one from the inner wall to a certain radial position and another one from the outer wall to another distinct radial position where the viscous force plays a major role. From the behaviour of u+ and y+, it is found that u+ = y+ up to certain radial location from both walls (see Figures 15 b-e). We defined viscous layer in the near wall region where azimuthal velocity varies linearly with distance from wall.

### 5.1.1 Spatial behaviour of VL in ATVF (Axisymmetric flow structure)

From the distribution of VL thickness along the Z direction in the r-Z plane at Θ = 0, 180 and at Re = 375 in the ATVF regime, it is found that the thickness of normalized VL (i.e., $\overline{VL}$) is minimum at the jet impingement (see Figure16) for both inner and outer walls and it increases from the jet impingement towards the flow separation, reaching its maximum value in the vicinity of flow separation region. As it approaches the flow separation region, a sudden decrease of $\overline{VL}$ is observed in a region denoted as the drop. This drop could be attributed to the emergence of a pair of secondary vortices in the flow separation region as reported in Akonur & Lueptow (2003) and Kristiawan, Jirout & Sobolík (2011). $\overline{VL}$ exhibits a similar drop for other Re in the flow separation region of ATVF (see Figure 17 and Figure 18). The drop observed in the $\overline{VL}$ provides vital information on the exact location of flow separation and its axial extent. The $\overline{VL}$ exhibits two identical peak values along the axial direction in the common flow separation region of two counter-rotating Taylor vortices (see Figure 16, Figure 17 and Figure 18).

The $\overline{VL}$ and normalized azimuthal wall shear stress (i.e., $\bar{\tau}_{r\theta}$) exhibits an inverse like relationship (see Figure 19 and Figure 20) from jet impingement until the vicinity of the flow separation region (see Figure 16) as seen and established for the open boundary flow and internal flow (Pope, 2000). For the inner and outer walls, a proportional like relationship (see Figure 19 and Figure 20) is observed in the common flow separation region of two counter-rotating Taylor vortices (see Figure 16). As far as we are aware, this behaviour of $\overline{VL}$ observed in the common flow separation region of two counter-rotating vortices has not been reported in the earlier studies.

### 5.1.2 Spatial behaviour of VL in the transition of ATVF into WVF (Non-axisymmetric flow structure)

The spatial behaviour of $\overline{VL}$ along the axial direction in the transition of ATVF into WVF flow regime for the outer and inner walls at Re = 475 in the r-Z planes at $\Theta$ = 0 and 180 is illustrated in Figure 21 and Figure 22 respectively. For $\Theta$ = 0 (see Figure 21), at the outer wall, the $\overline{VL}_o$ behaviour is similar to the one exhibited in ATVF regime (see Figure 16a) with the exception of distinct peak values where the greater peak is classified as global maximum and the lower peak is classified as local maximum (see Figure 21a). It has been shown in the earlier section, as the axisymmetric flow transforms into the non-axisymmetric flow (i.e., transition of ATVF into WVF), a strong periodic secondary flow appears in the flow separation region (see Figure 21b) of the outer wall and moves towards jet impingement region of the inner wall. The resultant flow then led to the global maximum of $\overline{VL}_o$ lying adjacent to the periodic secondary flow and the local maximum of $\overline{VL}_o$ is found where there is no secondary flow (see Figure 21a). As also mentioned earlier, the appearance of a strong periodic secondary flow (see Figure 7) results in lowering the strength of Taylor vortices and the occurrence of imbalance between two counter-rotating Taylor vortices (see Figure 9). This may be linked to the observed difference in the peak values of $\overline{VL}_o$ in the flow separation region.

For $\Theta$ = 0, the behaviour of the $\overline{VL}_i$ in the flow separation region of the inner wall (see Figure 21c) is similar to the one detailed in the inner wall of ATVF (see Figure 16c) with the exception of distinct peak values. Similar to the outer wall, the difference between the two peak values could be due to the appearance of imbalance between two counter-rotating Taylor vortices (see Figure 9) under the influence of strong periodic secondary flow (see Figure 7) in the flow separation region. Though the behaviour of $\overline{VL}_o$ at the jet impingement of the outer wall in this flow regime (see Figure 21a) is found identical to that of ATVF (see Figure 16a), the $\overline{VL}_i$ at the impingement of inner wall (i.e., in between point A and point C in Figure 21c) increases; unlike the $\overline{VL}_i$ of ATVF (see Figure 16c). The influence of strong periodic secondary flow might weaken the jet impingement at the inner wall there resulting in the appearance of a thicker $\overline{VL}_i$ at the inflow region of the inner wall. The distribution of $\overline{VL}_o$ and $\overline{VL}_i$ for $\Theta$ =180 (see Figure 22) is similar to that of $\Theta$ =0 (see Figure 21) with the exception of the shift in peak values due to the change in the direction of periodic secondary flow.

The behaviour of $\overline{VL}$ at the inner and outer walls as described above is also observed for other Re in this flow regime (see Figure 23).

### 5.1.3 Spatial behaviour of VL in WVF

The behaviour of $\overline{VL}$ along the axial direction at Re = 600 in the WVF for the outer and inner walls in the r-Z planes at $\Theta$ =0 and 180 has been illustrated in Figure 24 and Figure 25, respectively. For $\Theta$ =0 (see Figure 24), the $\overline{VL}_o$ in the outer wall along the axial direction in the WVF flow regime (see Figure 24a) is similar to that of the outer wall in the transition of ATVF into WVF flow regime (see Figure 21a).

For $\Theta$ =0, the behaviour of $\overline{VL}_i$ at the inner wall for WVF is similar to that of the transition flow (i.e., transition from ATVF into WVF) in the flow separation region (i.e., the appearance of global maximum and local maximum in $\overline{VL}_i$) (see Figure 21c and Figure 24c). However, a clear difference between the $\overline{VL}_i$ of transition flow (see Figure 21c) and WVF (see Figure 24c) is observed in the vicinity of inflow region of the inner wall (i.e., impingement region of the inner wall). As shown in the previous section, the $\overline{VL}_i$ increases in the inflow region of the inner wall in the transition flow (see Figure 21c) due to the emergence of weak jet impingement under the influence of strong periodic secondary flow. The periodic secondary flow becomes weaker in the WVF (see Figure 7). The weaker secondary flow cannot suppress the jet impingement region due to which $\overline{VL}_i$ becomes thinner in the inflow region of the inner wall in WVF. A larger difference between the two peak values (i.e., global maximum and local maximum in $\overline{VL}_i$) in the WVF in the inner wall (see Figure 24c) is also observed compared to that of the transition flow (see Figure 21c).

The distribution of $\overline{VL}_o$ and $\overline{VL}_i$ for $\Theta$ =180 is found similar to that of $\Theta$ =0 with the exception of a shift in the peak values in the flow separation region due to the change in the direction of periodic secondary flow (see Figure 25).

The behaviour of $\overline{VL}$ in the inner and outer walls as described above is also observed for other Re in this flow regime (see Figure 26).

### 5.2 Variation of the spatial behaviour of VL with Re

In this section, the variation of $\overline{VL}$ along the axial direction with Re in the r-Z plane at $\Theta$=0 is further examined for both the axisymmetric and non-axisymmetric flows. In ATVF (see Figure 16 and Figure 17), transition flow (see Figure 23) and WVF ( see Figure 26), the variation of $\overline{VL}$ with Re at the jet impingement, flow separation and drop for the outer and inner walls has been summarized in Figure 27a and Figure 27b, respectively. The variation of the ratio between the two peak values across the flow separation with Re in the ATVF, the transition between ATVF into WVF and WVF regimes is illustrated in Figure 28.

### 5.2.1 Variation of the spatial behaviour of VL with Re in the ATVF (Axisymmetric flow structure)

For the case of the outer wall, the $\overline{VL}_o$ exhibits a prominent dependence on Re at the jet impingement, flow separation and drop (see Figure 27a). At the jet impingement, a monotonic decrease is observed in the $\overline{VL}_o$ with the increase in Re. A gradual increase in $\overline{VL}_o$ is observed at the drop up to Re = 275 beyond which it decreases gradually until Re = 425 (see Figure 27a). At the flow separation, $\overline{VL}_o$ decreases gradually till Re = 225; from Re =225 to 275, $\overline{VL}_o$ starts to increase and reaching its local maximum at Re = 275 beyond which it exhibits a gradual drop until Re = 425. In can be inferred that at the flow separation and drop, $\overline{VL}_o$ exhibits significant change as Re is increased beyond 225 (see Figure 27a). In the earlier section, it was mentioned that at Re = 225, NNW is found to be 1.60 with 10 cells of Taylor vortices and NNW is found to be 2.00 with 8 cells of Taylor vortices as Re is increased to 275 (see Figure 10 and Figure 11). Thus, the occurrence of a sudden change in $\overline{VL}_o$ from Re =225 to 275 in the flow separation region of the outer wall could be due to the occurrence of a secondary form of instability which may eventually contribute to the transformation of NNW from 1.60 (i.e., 10 cells of Taylor vortices) to NNW = 2.00 (i.e., 8 cells of Taylor vortices) (see Figure 10 and Figure 11). The $\overline{VL}_i$ at the flow separation and drop is found nearly independent of Re in the inner wall up to Re =425 (see Figure 27b). However, increase in Re results in slightly decreasing trend of $\overline{VL}_i$ at the jet impingement of the inner wall (see Figure 27b).

The ratio between two peak values of $\overline{VL}$ across the flow separation region of the inner and outer wall is nearly identical up to Re =425 (see Figure 28) which could be attributed to the existence of equilibrium between counter-rotating Taylor vortices (see Figure 9).

### 5.2.2 Variation of the spatial behaviour of VL with Re in the transition of ATVF into WVF and WVF

For the outer wall, there is a sudden increase in $\overline{VL}_o$ at the flow separation and drop during the initial phase of the transition flow from Re = 425 until Re = 475 following which a descent is observed till the end of the transition flow (i.e., Re = 575) (see Figure 27a). It follows a steeper descent in the WVF with an inflection point at the beginning of the WVF (Re = 575) (see Figure 27a). The $\overline{VL}_o$ in the jet impingement also exhibits a very slight increase at the beginning of the transition flow (i.e., Re = 425) and a gradual decrease until the end of the WVF (i.e., Re = 650). The change in $\overline{VL}_o$ in the jet impingement is not as significant as that of the flow separation and drop in the transition flow (see Figure 27a). The ratio between the two peak values of $\overline{VL}_o$ starts to increase at the beginning of the transition flow, reaches its maximum value at Re = 475 and decreases from Re = 475 until the end of transition flow (i.e., Re = 575). In the primary phase of WVF till Re = 600, it is observed to be increasing following which a descent is observed (see Figure 28).

In the inner wall (see Figure 27b), the variation of $\overline{VL}_i$ at the flow separation, jet impingement and drop does not exhibit any significant change in the transition flow with the increase in Re (i.e., 425 < Re < 575). In the beginning of WVF from Re = 575, there is a sudden jump observed in the $\overline{VL}_i$ at the flow separation and drop whereas at the jet impingement, it continues its steady descent from the end of the transition flow(see Figure 27b). At the beginning of transition flow from Re = 425, the ratio between the two peak values of $\overline{VL}_i$ increases and reaches its local maximum at Re = 475 beyond which it decreases until the end of the transition flow (see Figure 28). There is a sudden boost observed in the ratio between peak values of $\overline{VL}_i$ at the starting of WVF from Re = 575 till Re = 650 (see Figure 28). The change in the ratio between peak values of $\overline{VL}_i$ at the beginning of WVF is far more significant than that of the transition flow (see Figure 28). In the transition flow, the change in the ratio between two peak values of $\overline{VL}$ is found to be significant at the outer wall whereas in the WVF, the change is found to be significant at the inner wall.

## 5.2.3 The source of non-axisymmetric disturbances, its propagation and its influence in the behaviour of VL thickness

### 5.2.3 (a) Source of non-axisymmetric disturbances

As described in the previous section, from the study of near–wall region, the $\overline{VL}$ at the jet impingement region at the outer and inner walls does not exhibit any significant change in the transition flow and WVF (see Figure 27). Thus, it is obvious that the jet impingement region of the inner and outer walls does not have any significant contribution to the occurrence of non-axisymmetric flow structure (i.e., WVF flow regime) in wide gap problems which contrasts with the findings of Snyder & Lambert (1965), Fazel & Booz (1984), Marcus (1984), Dumont et al. (2002) and Akonur & Lueptow (2003) for small gap problems. In the transition flow, the flow separation region of the outer wall undergoes significant changes unlike the flow separation region of the inner wall (see Figure 27 and Figure 28). Hence, the flow separation region of the outer wall may be deemed to play a dominant role in the appearance of the non-axisymmetric flow structure in the transition flow. It has been described in the earlier section that as Re is increased beyond 425, flow structure transforms into non-axisymmetric flow structure with a strong azimuthal wave in the inflow region and comparatively weak azimuthal wave in the outflow region (see Figure 8). From the flow structure shown in Figure 6, the flow separation region lies at the inflow region for the outer wall and outflow region for the inner wall. Thus, this provides a strong evidence that the flow behaviour in the flow separation region of the outer wall plays a significant role in the appearance of non-axisymmetric flow structure and azimuthal wave in wide gap problems which is partially aligned with the study of small gap problems by Kristiawan, Jirout & Sobolík (2011) where the source of non-axisymmetric disturbances was considered to be at the flow separation region of the inner and outer wall.

### 5.2.3 (b) Propagation of non-axisymmetric disturbances in a form of periodic secondary flow

From the behaviour of $\overline{VL_o}$, it can be suggested that as Re is increased beyond 425, the flow separation region of the outer wall may experience the onset of sudden instability resulting in an unstable inflow region. The instability in the inflow region magnifies over time and results in the propagation of periodic non-axisymmetric disturbances in the form of strong periodic secondary flow (see Figure 6 and Figure 7). Eventually, a strong azimuthal wave in the inflow region and a comparatively weaker azimuthal wave in the outflow region is observed (see Figure 9).

### 5.2.3(c) The influence of periodic secondary flow in NNW and torque in the transition flow

As shown in the earlier section, during the transition flow, a sudden increase in NNW is observed (see Figure 12). In the same flow regime, a maximum of 11% torque reduction is observed vis-a-vis ATVF (see Figure 13 and Figure 14). It was also indicated that the strong periodic secondary flow (see Figure 7) results in the sudden increase in NNW (see Figure 11) and substantial decrease in the strength of the two counter-rotating Taylor vortices (see Figure 9) in the transition flow. From the behaviour of $\overline{VL}$ in the flow separation region of the transition flow, it is well suggested that the appearance of the strong periodic secondary flow contributes to the appearance of thicker $\overline{VL}$ (see Figure 27) by lowering the strength of the two counter-rotating Taylor vortices. Thus the observed torque reduction in this flow regime is directly related to the appearance of a thicker VL which is aligned with the findings of Mangiavacchi & Akhavan (1992), Laadhari, Skandaji & Morel (1994), Baron and Quadrio (1995), Choi & Graham (1998) and Choi & Clayton (2001).

### Conclusion

DNS of a Taylor Couette flow has been conducted for infinite length of cylinders with a radius ratio of 0.5 and Reynolds number (Re) ranging from 60 to 650. The objective is to investigate the source of non-axisymmetric disturbances and its propagation mechanism for wide gap problems to better understand the transition of axisymmetric Taylor vortex flow into the non-axisymmetric flow (i.e., wavy vortex flow). The use of four wavelengths fluid column in the present study has uncovered detail physical phenomenon involved in the non-axisymmetric flow which was not possible via a single wavelength fluid column which has been generally employed in the numerical study of Taylor Couette flow. The results show that the flow structure is axisymmetric up to Re = 425; beyond which it transforms into a non-axisymmetric flow. From the spatial behaviour of VL thickness along the axial direction in the r-Z plane, the VL thickness is found minimum in the jet impingement region and reaches its peak value near the flow separation region. A drop in the VL thickness is observed across the flow separation region between two peak values. These peak values of VL thickness are nearly identical in the axisymmetric flow but a clear difference is observed in the non-axisymmetric flow. From the behaviour of VL thickness with Re in the transition flow, it is found that flow *separation* and jet impingement regions of the inner wall and jet impingement region of outer wall do not exhibit any significant change. However, a sudden increase in the VL thickness is observed in the flow *separation* region of the outer wall in the transition flow. This suggests that the flow separation region of the outer wall is likely the source of the appearance of non-axisymmetric disturbances in the transition flow in TCF of wide gap problems. These non-axisymmetric disturbances develop in the form of periodic secondary flow. The strength of this periodic secondary flow becomes stronger in the middle of transition flow but starts to die out at the end of transition and the beginning of wavy vortex flow. In addition, this also contributes to the sudden increase in the natural wavelength and decrease in the strength of Taylor vortices which leads to the appearance of a thicker VL in the inner and outer wall. This in turns contributes to the reduction of torque in the transition flow vis-à-vis the axisymmetric Taylor vortex flow. The appearance of periodic secondary flow results in the occurrence of azimuthal wave in the inflow and outflow regions which contributes to the occurrence of unequal strength between two counter-rotating Taylor vortices in the transition and wavy vortex flow. The unequal strength of counter-rotating Taylor vortices observed in the transition and wavy vortex flow leads to the emergence of the distinct difference between two peak values of VL thickness across the flow separation region of the inner and outer walls. The findings obtained in this study may help in providing further insights on the complexities involved in the transition states of Taylor Couette flow for wide gap problems.

## Acknowledgements

The authors acknowledge the NUS Research Scholarship, the advice from Dr. T.T. Lim on the content of this manuscript and computational support from Dr. Y. J. Lee. The numerical computation has been performed on the resources of High Performance Computation (HPC) of the National University of Singapore and National Supercomputing Centre, Singapore.

## References

Abshagen, J., Lopez, J., Marques, F., & Pfister, G., Bursting dynamics due to a homoclinic cascade in Taylor–Couette flow. Journal of Fluid Mechanics, 613, 357-384 (2008).

Abshagen, J & von Stamm, J & Heise, M & Will, Ch & Pfister, Gerd., Localized modulation of rotating waves in Taylor-Couette flow. Physical Review E. 85. 56307 (2012).

Akonur, A. & Lueptow, R. M., Three-dimensional velocity field for wavy Taylor– Couette flow. Physics of Fluids, 15(4) (2003).

Baron, A., Quadrio, M., Turbulent Drag Reduction by Spanwise Wall Oscillations. Applied Scientific Research, 311–326 (1996).

Blennerhassett, P. J. & Hall, P., Centrifugal Instabilities of Circumferential Flows in Finite Cylinders: Linear Theory. Proc. Roy. Soc. London, A, 365(1721), 191-207 (1979).

Bilson, M., & Bremhorst, K. (2007). Direct numerical simulation of turbulent Taylor–Couette flow. Journal of Fluid Mechanics, 579, 227-270.

Chandrasekhar, S., The Stability of Viscous Flow between Rotating Cylinders in the Presence of a Magnetic Field. Proceedings of the Royal Society of London. Series A, Mathematical and Physical Sciences, 216(1126), 293-309 (1953).
.
Chandrasekhar, S., The Stability of Viscous Flow between Rotating Cylinders. Royal Society, 246(1246) (1958).

Choi, K. S., & Graham, M., Drag reduction of turbulent pipe flows by circular-wall oscillation. Physics of Fluids, 10(1), 7–9 (1998).

Choi, K., & Clayton, B. R. , The mechanism of turbulent drag reduction with wall oscillation. International Journal of Heat and Fluid Flow, 22, 1–9 (2001).

Cole, J. A. , Taylor vortex instability and annulus length effects. Nature, 252, 688 – 689 (1974).

Coles, D. , Transition in circular Couette flow. Journal of Fluid Mechanics, 21, 385-425 (1965).

Davey, A., The growth of Taylor vortices in flow between rotating cylinders. Journal of Fluid Mechanics, 14(3), 336-368 (1962).

Dessup, T., Tuckerman, L. S., Wesfreid, J. E., Barkley, D., & Willis, A. P. (2018). Self-sustaining process in Taylor-Couette flow. Physical Review Fluids, 3(12).

Dong, S. (2007). Direct numerical simulation of turbulent Taylor–Couette flow. Journal of Fluid Mechanics, 587, 373-393.

Donnelly, R. J., Experiments on the Stability of Viscous Flow between Rotating Cylinders. I. Torque. Proceedings of the Royal Society of London. Series A, Mathematical and Physical, 246(1246), 312-325 (1958).

Donnelly, R. J. & Fultz, D. , Experiments on the Stability of Viscous Flow Between Rotating Cylinders. II. Visual Observation. Proceedings of the Royal Society of London. Series A, Mathematical and Physical Sciences, 258, 101-123 (1960).

Donnelly, R. J. & Simon, N. J., An empirical torque relation for supercritical flow between rotating cylinders. Journal of Fluid Mechanics, 7(3), 401-418 (1960).

Dumont, E., Fayolle, F., Sobolík, V. & Legrand, J. , Wall shear rate in the Taylor–Couette–Poiseuille flow at low axial Reynolds number. International Journal of Heat and Mass Transfer, 45(3), 679-689 (2002).

Fasel, H. & Booz, O. , Numerical investigation of supercritical Taylor-vortex flow for a wide gap. J. Fluid Mec., 138, 21-52 (1984).

Gao, X., Kong, B. & Vigil, R. D., CFD simulation of bubbly turbulent Tayor–Couette flow. Chinese Journal of Chemical Engineering., 26(6), 719-727 (2016).

Gorman, M. & Swinney, H. L., Visual Observation of the Second Characteristic Mode in a Quasiperiodic Flow. Physic Rev. Lett., 43(24) (1979)..

Hall, P., Centrifugal Instabilities of Circumferential Flows in Finite Cylinders: Nonlinear Theory. Proc. Roy. Soc. London A, 372(1750), 1794-1797 (1980).

Heise, M, J. Abshagen, D. Küter, K. Hochstrate, G. Pfister, and Hoffman, C., Localized spirals in Taylor-Couette flow. Phys. Rev. E 77, 026202 (2008).

Huisman, G.S., Scharnowski, S., Cierpka, C., Kahler, C.J., Lohse, D & Chao, S., Logarithmic boundary layers in highly turbulent Taylor-Couette flow. Phys. Rev. Lett., 110, 264501 (2013).

Jeng, J., & Zhu, K. Q. (2010). Numerical simulation of Taylor Couette flow of Bingham fluids. Journal of Non-Newtonian Fluid Mechanics, 165(19–20), 1161–1170.

Kristiawan, M., Jirout, T., & Sobolík, V. (2011). Components of wall shear rate in wavy Taylor-Couette flow. Experimental Thermal and Fluid Science, 35(7), 1304–1312.

Kataoka, K., Doi, H. & Komai, T., Heat/mass transfer in Taylor vortex flow with constant axial flow rates. Experimental Thermal and Fluid Science, 20, 57-63 (1977).

Laadhari, F., Skandaji, L., & Morel, R., Turbulence reduction in a boundary layer by a local spanwise oscillating surface. Physics of Fluids, 6(10), 3218–3220 (1994).

Lim, T. T., Chew, Y.T. & Xiao Q.,  A new flow regime in a Taylor–Couette flow. Physics of Fluids, 10(12) (1998).

Lim, T. T. & Tan, K. S., A note on power-law scaling in a Taylor–Couette flow. Physics of Fluids, 16(140) (2004).

Liao, C. B., Jane, S. J., & Young, D. L. (1999). Numerical simulation of three-dimensional Couette-Taylor flows. *International Journal for Numerical Methods in Fluids*, *29*(7), 827–847.

López, Mańıa, J. and Francisco, M., "Dynamics of axially localized states in Taylor-Couette flows." Physical Review. E, Statistical, nonlinear, and soft matter physics , 91(5) (2015).

Manneville, P. & Czarny, O., Aspect-ratio dependence of transient Taylor vortices close to threshold. Theor. Comput. Fluid Dyn. 23: 15 (2009).

Mangiavacchi, N., Jung, W. J., & Akhavan, R., Suppression of turbulence in wall-bounded flows by high-frequency spanwise oscillations. Physics of Fluids A, 4(8), 1605–1607 (1992).

Martinand, D., Serre, E., & Lueptow, R. M. (2014). Mechanisms for the transition to waviness for Taylor vortices. Physics of Fluids, 26(9).

Martinand, D., Serre, E., & Lueptow, R. M. (2017). Linear and weakly nonlinear analyses of cylindrical Couette flow with axial and radial flows. Journal of Fluid Mechanics, 824, 438–476.

Marcus, P. S., Simulation of Taylor-Couette flow. Part 2. Numerical results for wavy-vortex flow with one travelling wave. Journal of Fluid Mechanics, 146, 65-113 (1984).

Ng, J. H., Jaiman, R. K., & Lim, T. T. (2018). Interaction dynamics of longitudinal corrugations in Taylor-Couette flows. Physics of Fluids, 30(9).

Pope, S., Turbulent Flows. Cambridge: Cambridge University Press (2000).

Pirrò, D., & Quadrio, M. (2008). Direct numerical simulation of turbulent Taylor-Couette flow. European Journal of Mechanics, B/Fluids, 27(5), 552–566.

Rodolfo, O.E. , Cierpka, Verzicco,R.,C.J., Grossmann,S., & Detlef, L., Boundary layer dynamics at the transition between the classical and the ultimate regime of Taylor-Couette flow. Physics of Fluids., 26, 015114 (2014).

Singh, H., Suazo, C. A and Liné, A., Log law of the wall revisited in Taylor-Couette flows at intermediate Reynolds numbers. Phys. Rev. E 94, 053120 (2013).

Snyder, H. A. & Lambert, R. B., Harmonic generation in Taylor vortices between rotating cylinders. Journal of Fluid Mechanics, 26(3), 545-562 (1965).

Stuart, J. T., On the non-linear mechanics of hydrodynamic stability. Journal of Fluid Mechanics, 4, 1 (1957).

Stuart, J. T., Proceedings of the 10th International Congress of Applied Mechanics. Elsevier Publishing Company. Amsterdam (1960).

Sobolik, V., Jirout, T., Havlica, J. & Kristiawan, M., Wall Shear Rates in Taylor Vortex Flow. Journal of Applied Fluid Mechanics, 4(2), 25-31 (2011).

Sobolik, V., Benabes, B.& Cognet, G., Study of Taylor-Couette flow using a three-segment electro-diffusion probe. Journal of Applied Electrochemistry, 15(1), 441–449 (1995).

Sparrow, E. M., Munro, W. D. & Jonsson, V. K., Instability of the flow between rotating cylinders: the wide-gap problem. Journal of Fluid Mechanics, 20(1), 35-46 (1964).

Synge, J. L., On The Stability Of A Viscous Liquid Between Rotating Coaxial Cylinders. Proceedings of the Royal Society of London. Series A, Mathematical and Physical, 167, 250-256 (1938).

Taylor, G. I., Stability of a Viscous Liquid Contained between Two Rotating Cylinders. hilosophical Transactions of the Royal Society of London. Series A, Containing Papers, 223, 289-343 (1923).

Walden, R. W. & Donnelly, R. J., Reemergent Order of Chaotic Circular Couette Flow. Phys. Rev. Lett., 42, 301 (1979).

Watson, J., On the non-linear mechanics of wave disturbances in stable and unstable parallel flows Part 2. The development of a solution for plane Poiseuille flow and for plane Couette flow. Journal of Fluid Mechanics, 9(3), 371-389 (1960).

Wu, J. Z., Ma, H. Y., & Zhou, De, M. Vorticity and vortex dynamics. Vorticity and Vortex Dynamics

(2006).

**Figures:**

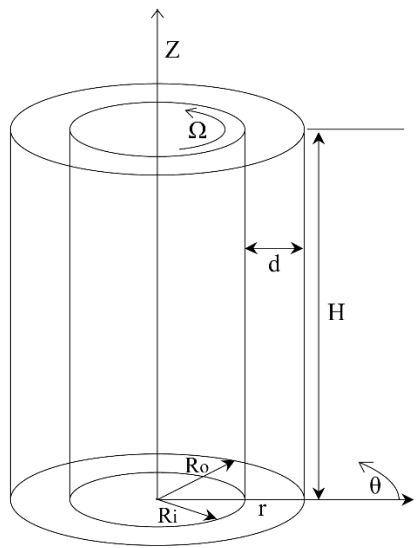

Figure 1: A schematic diagram of Taylor-Couette flow configuration.

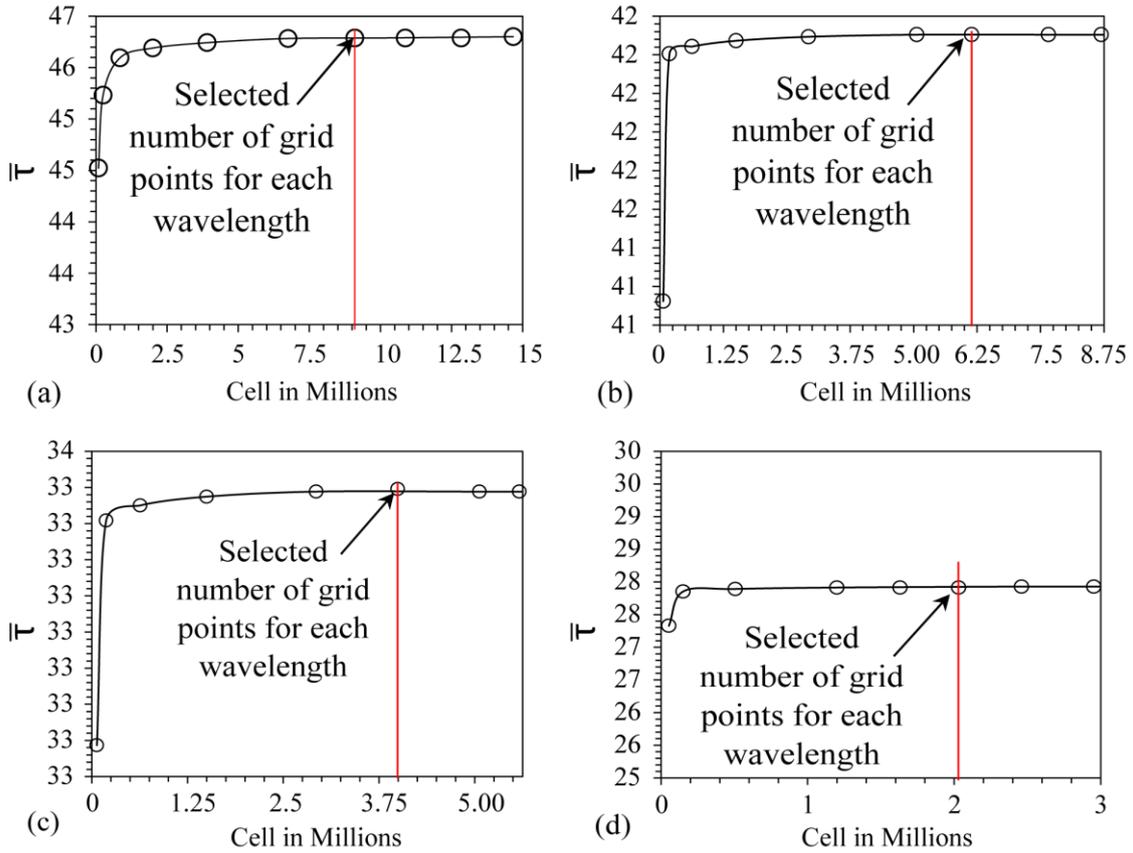

Figure 2: Grid convergence test for Re ranges between 60 to 650 (a) Grid independency for 650 (475≤Re≤650) (b) Grid independency for 475 (275≤Re≤475 )(c) Grid independencies for Re 275 (175≤Re≤275)(d) Grid independencies for 175 (60≤Re≤175).

$\bar{\tau}$

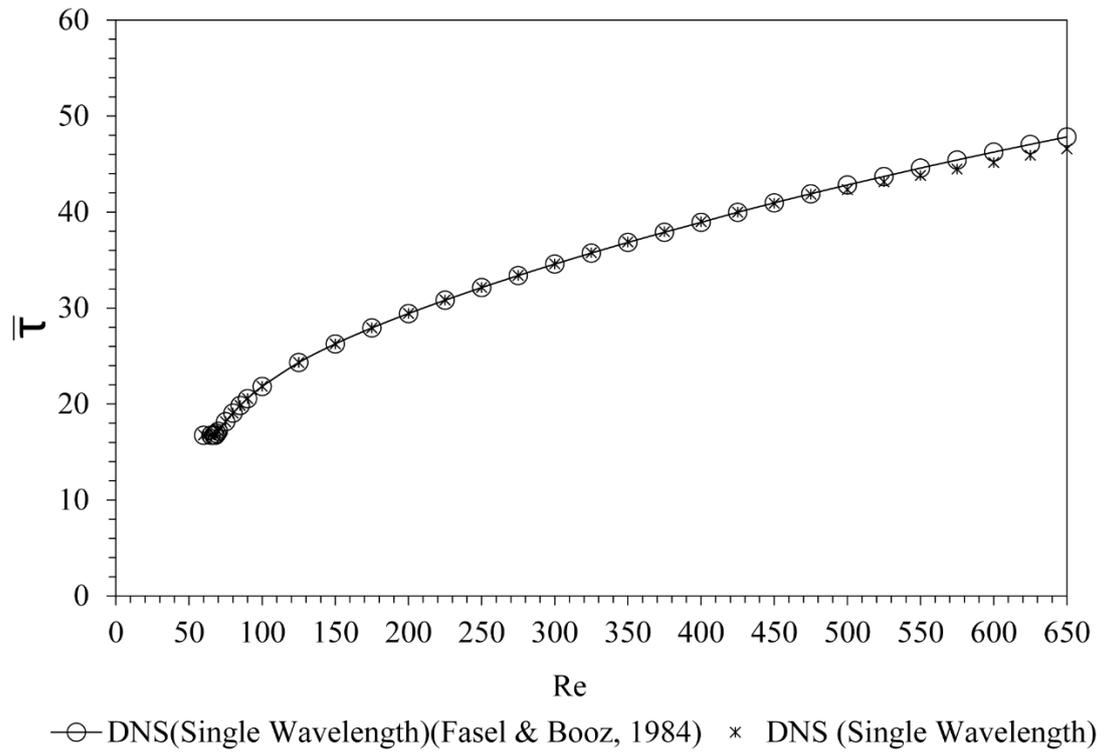

Figure 3: Comparison of normalized torque ($\bar{\tau}$) obtained in present DNS against Fasel & Booz (1984) for a single wavelength of the fluid column.

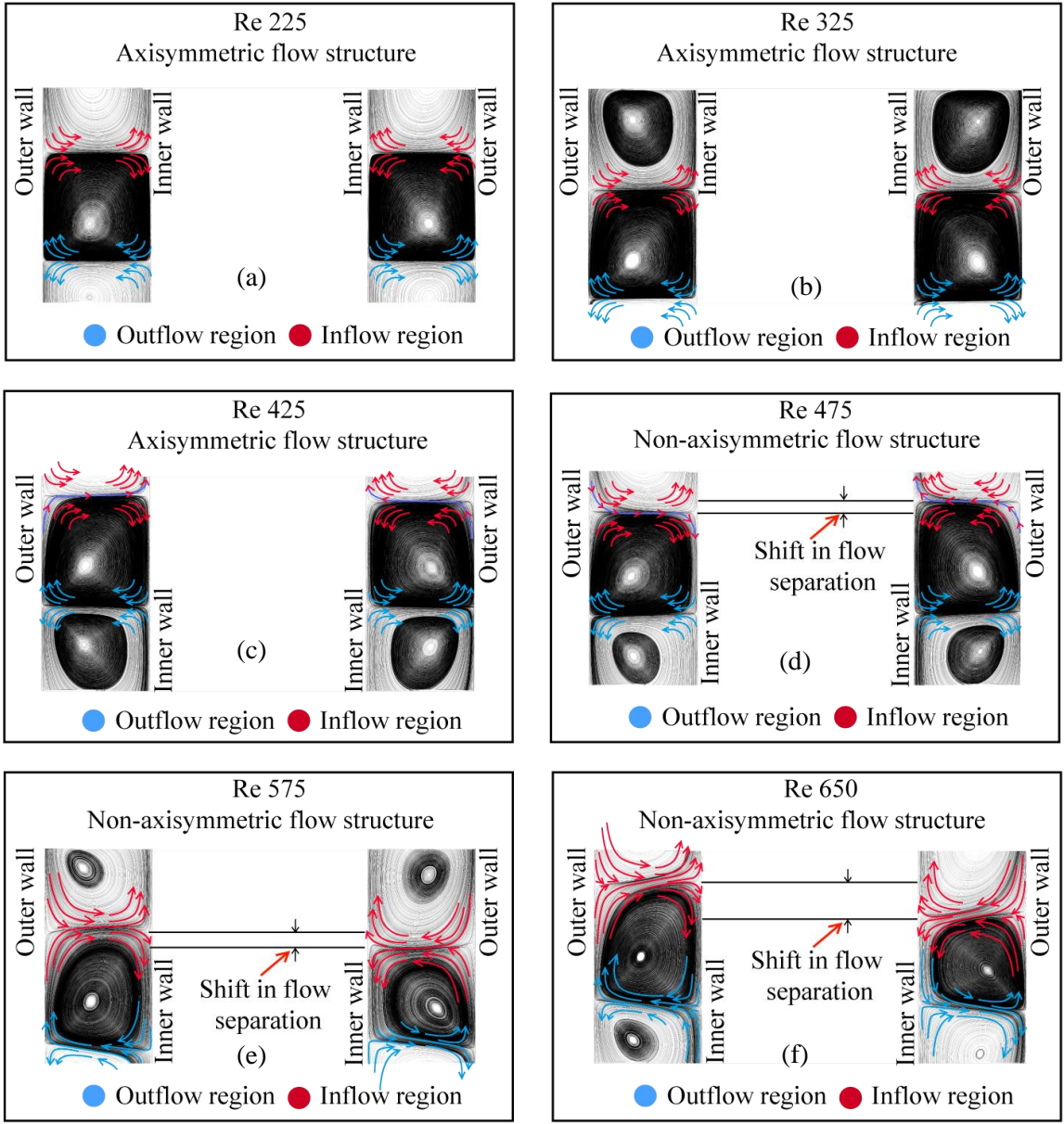

Figure 4: Streamline patterns associated with flow structure variation in TCF for radius ratio 0.5, single wavelength fluid column.

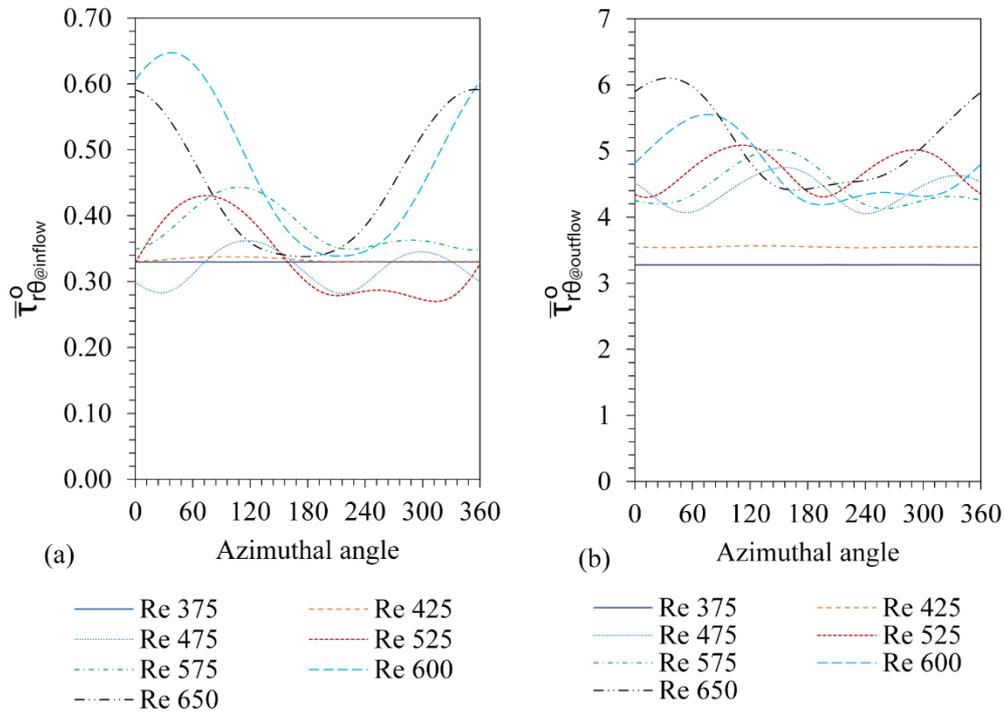

Figure 5: The distribution of normalized azimuthal shear stress ($\bar{\tau}_{r\theta}^{o}$) along the azimuthal direction at a fixed axial position of the outer wall (a) the distribution of $\bar{\tau}_{r\theta}^{o}$ at inflow region (b) distribution of $\bar{\tau}_{r\theta}^{o}$ at outflow region.

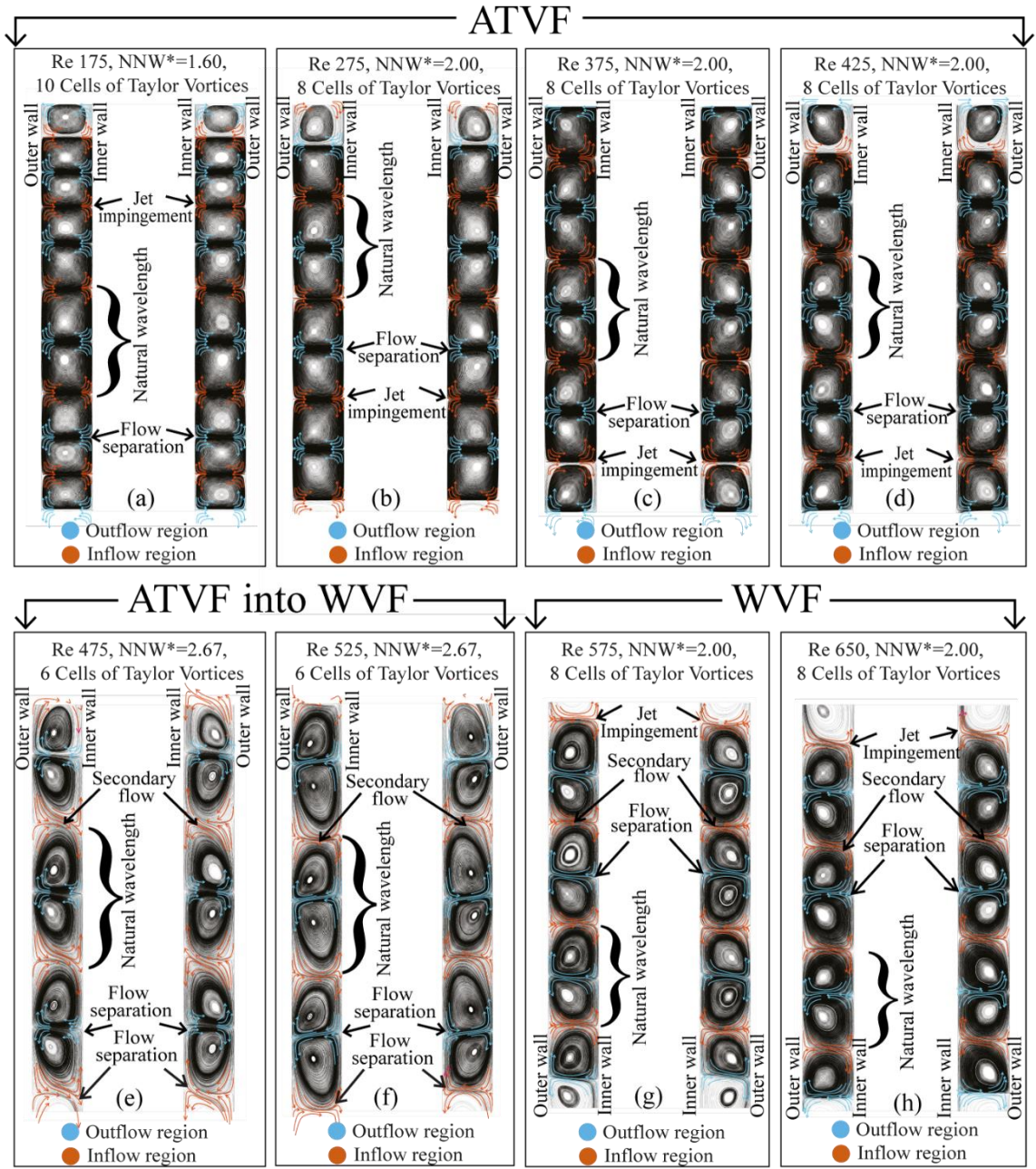

Figure. 6: Streamline patterns associated with flow structure variation in Taylor Couette flow for radius ratio 0.5, four wavelength and infinite aspect ratio(a)-(d)ATVF (e) transition of ATVF into WVF, (f) WVF.

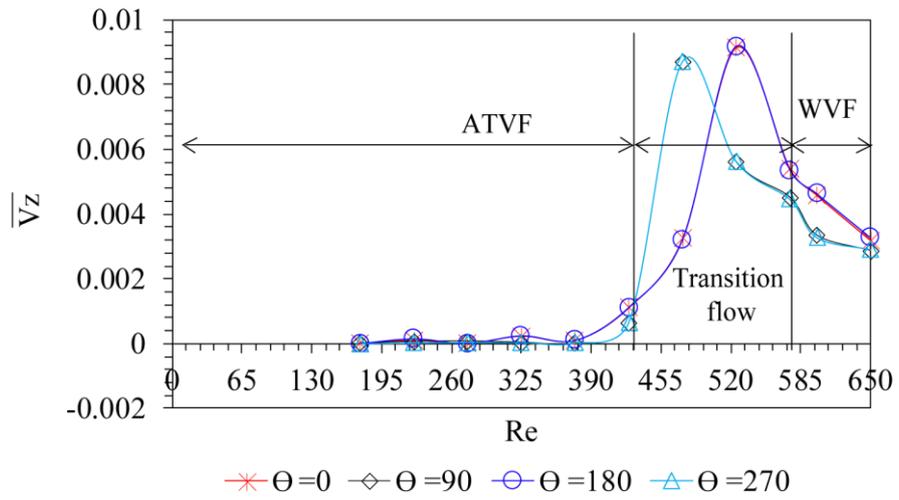

Figure 7: The variation of the magnitude of area average normalized axial velocity ($\bar{V}_z$) in the r-Z plane at $\Theta$ =0, 90, 180 and 270.

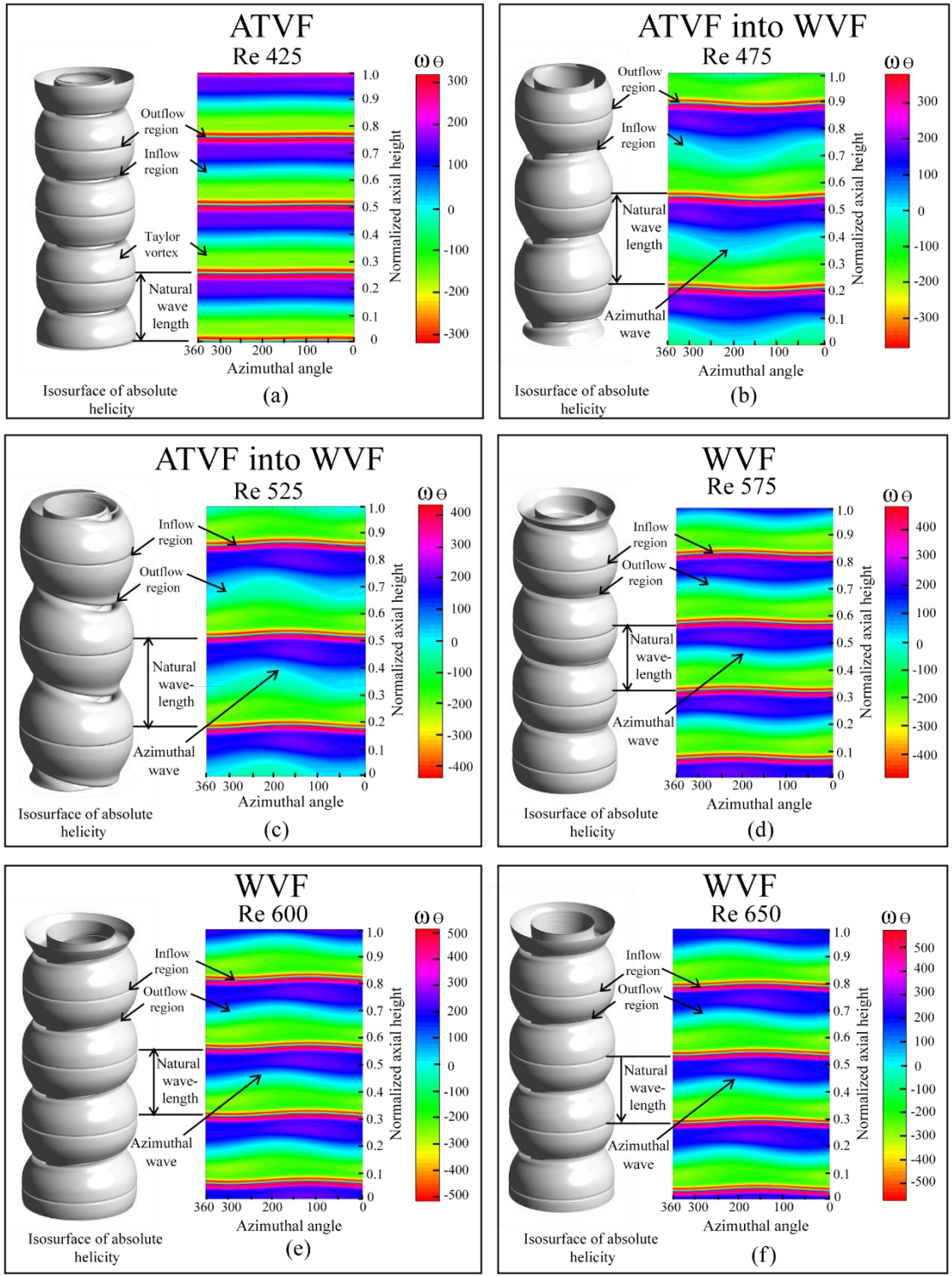

Figure 8: The azimuthal vorticity ($\omega_\theta$) contour around 360 degrees at the middle of annular gap (a) $\omega_\theta$ contour in the ATVF (b) –(c) $\omega_\theta$ contour transition in the ATVF into WVF (d) –(f) $\omega_\theta$ contour in the WVF.

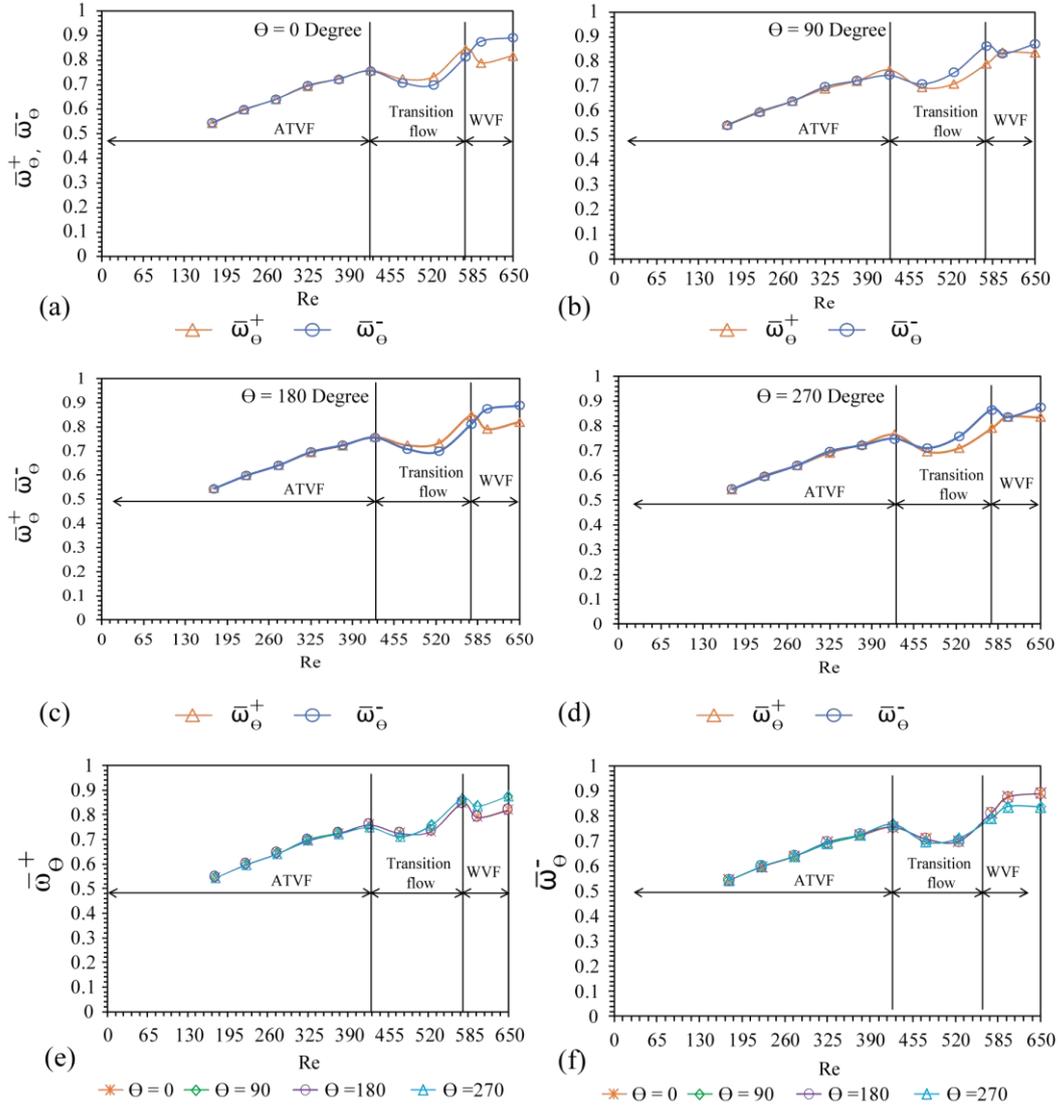

Figure 9: The variation of area average normalized azimuthal vorticity $\bar{\omega}_\Theta^+$ (clockwise) and $\bar{\omega}_\Theta^-$ (counter clockwise) with Re in ATVF, transition of ATVF into WVF and WVF flow regimes in the r-Z plane (a)-(d) variation of $\bar{\omega}_\Theta^+$ (clockwise) and $\bar{\omega}_\Theta^-$ (counter clockwise) with Re at Θ=0, 90, 180 and 270, respectively. (a) variation of $\bar{\omega}_\Theta^+$ with Re at Θ=0, 90, 180 and 270 (b) ) variation of $\bar{\omega}_\Theta^-$ with Re at Θ=0, 90, 180 and 270

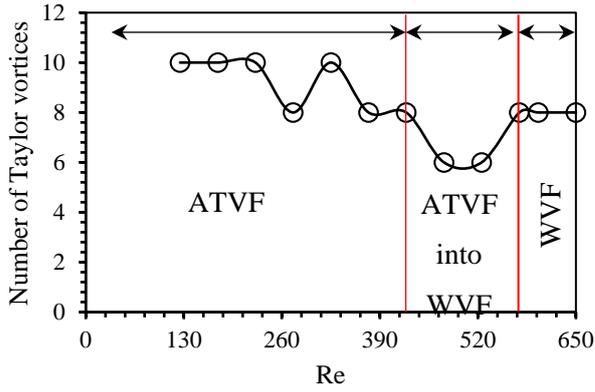
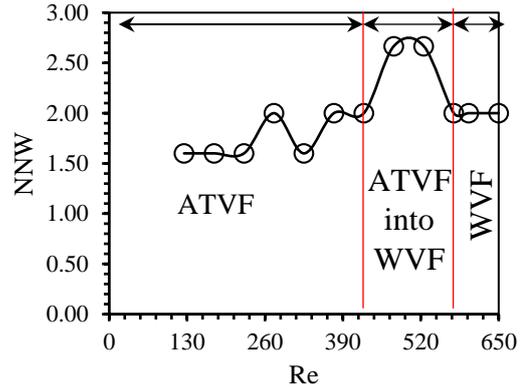

Figure 10: The variation of number of Taylor vortices with Re.

Figure 11: The variation of normalized natural wavelength with Re.

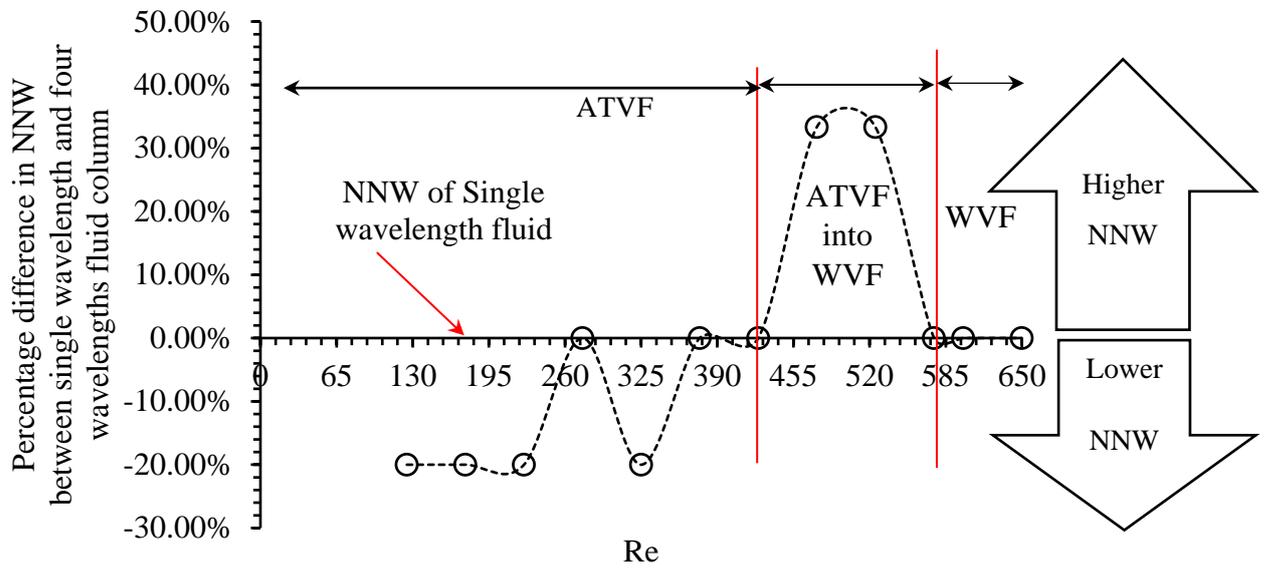

Figure 12: The percentage difference in NNW between single and four wavelengths fluid column.

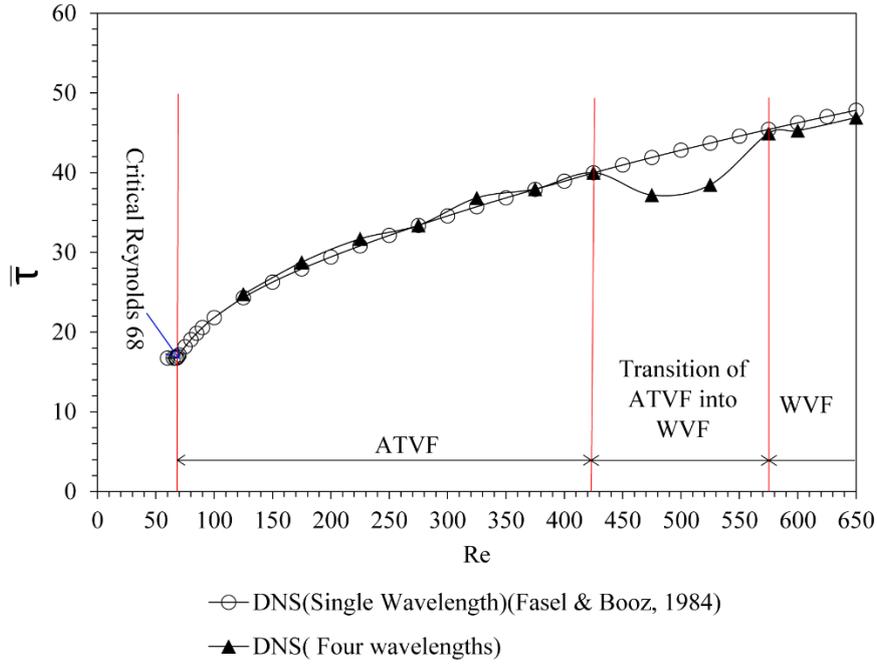

Figure 13: Comparison of normalized torque ($\bar{\tau}$) obtained in present DNS (Four wavelengths of the fluid column) against the study of Fazel & Booz (1984) for a single wavelength of the fluid column.

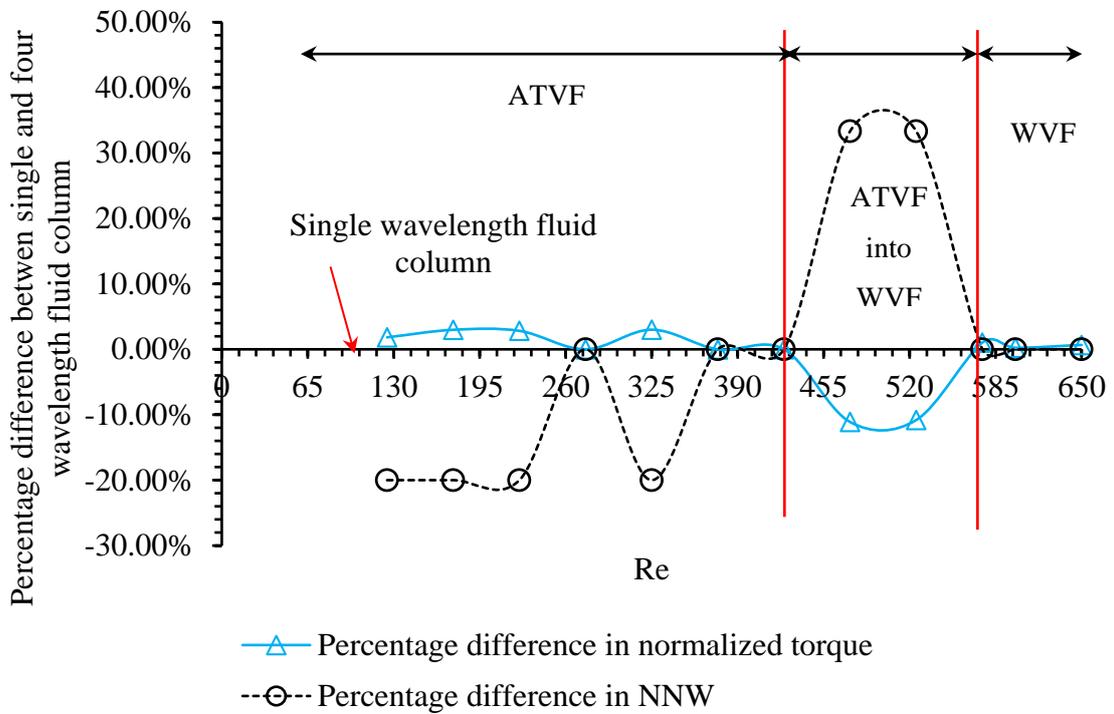

Figure 14: The percentage difference of normalized torque and NNW obtained in four wavelengths fluid column in comparison to the single wavelength fluid column.

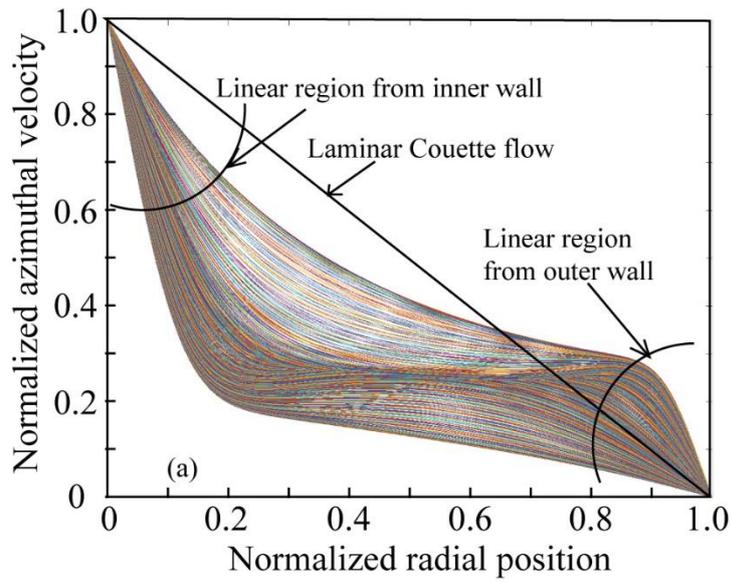

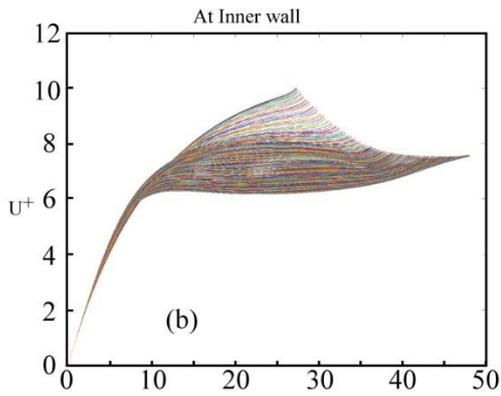
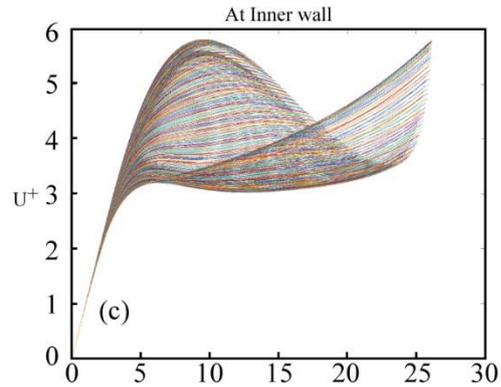

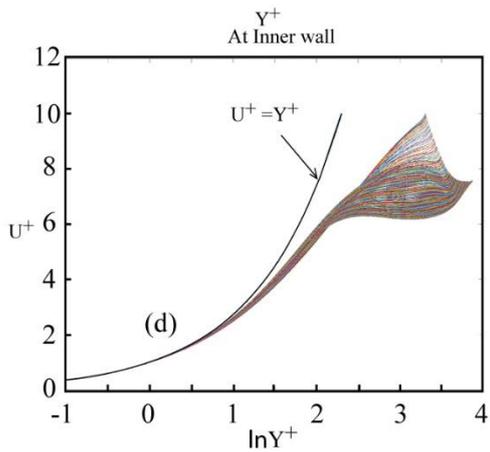
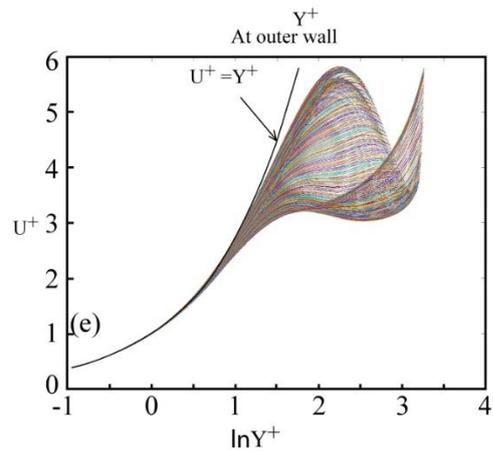

Figure 15: The variation of azimuthal velocity, U+ and U+ at different axial location in the r-Z plane of Θ=0 and Re = 425(a) Azimuthal velocity profile (b) the relation between U+ and U+ at the inner wall(c) the relation between U+ and Y+ at the outer wall (d) relation between U+ and lnY+ at the inner wall (e ) relation between U+ and lnY+ at the outer wall.

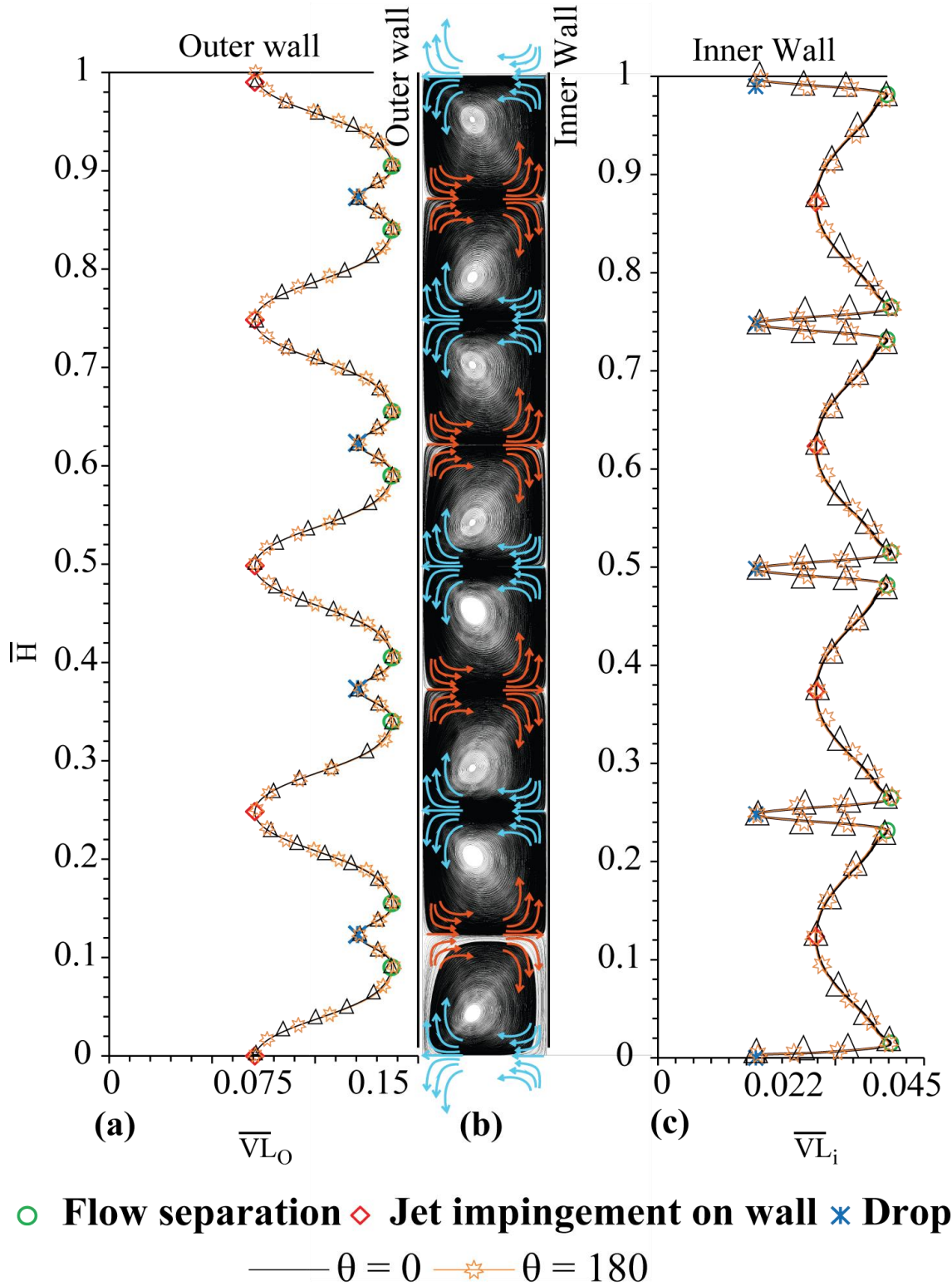

Figure 16: The distribution of normalized VL thickness for axisymmetric flow structure along the axial direction for inner and outer wall for Re 375 in the r-Z plane at $\Theta=0$ and 180 (a) $\overline{VL}_o$ distribution at outer wall (b) Streamlines associated with vortex structure (c) $\overline{VL}_i$ distribution at inner wall.

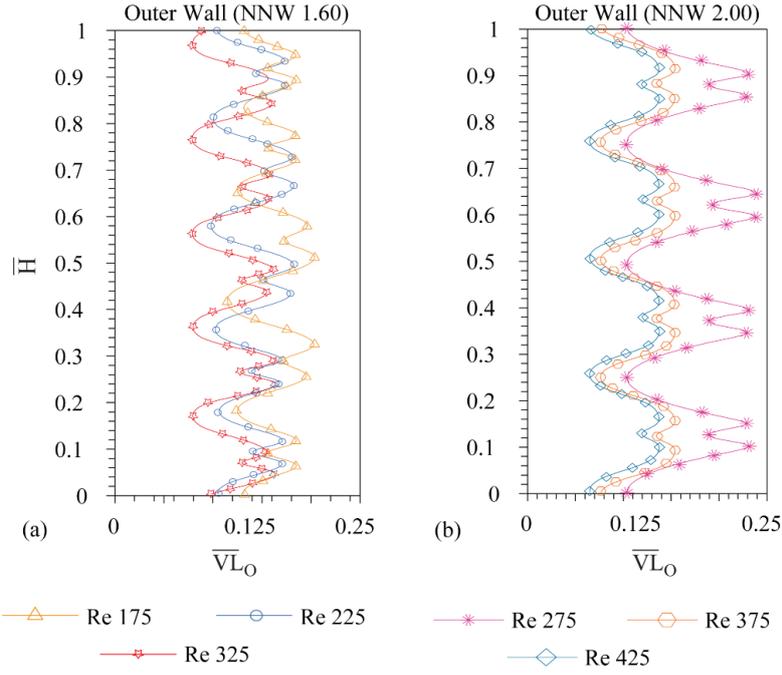

Figure 17: The spatial behaviour of $\overline{VL}_o$ along the axial direction for axisymmetric flow structure for the outer wall in the r-Z plane at $\Theta=0$ (a) $\overline{VL}_o$ for 10 cells of Taylor Vortices and normalized natural wavelength 1.60 (b) $\overline{VL}_o$ for 8 cells of Taylor Vortices and normalized natural wavelength 2.00.

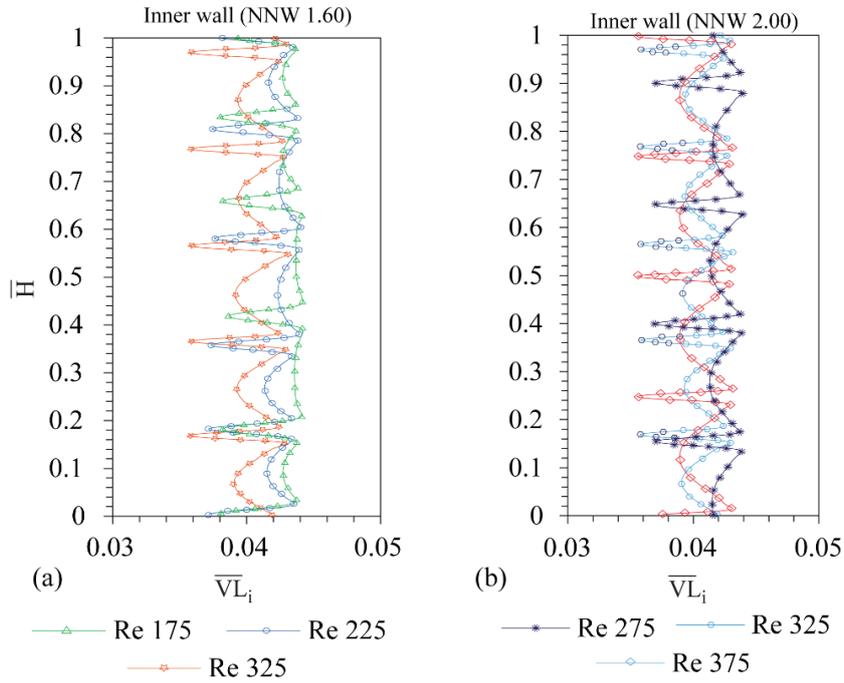

Figure 18: The spatial behaviour of $\overline{VL}_i$ along the axial direction for axisymmetric flow structure for the inner wall in r-Z plane at $\Theta=0$ (a) $\overline{VL}_i$ for 10 cells of Taylor Vortices and normalized natural wavelength 1.60 (b) $\overline{VL}_i$ distribution for 8 cells of Taylor Vortices and normalized natural wavelength 2.00.

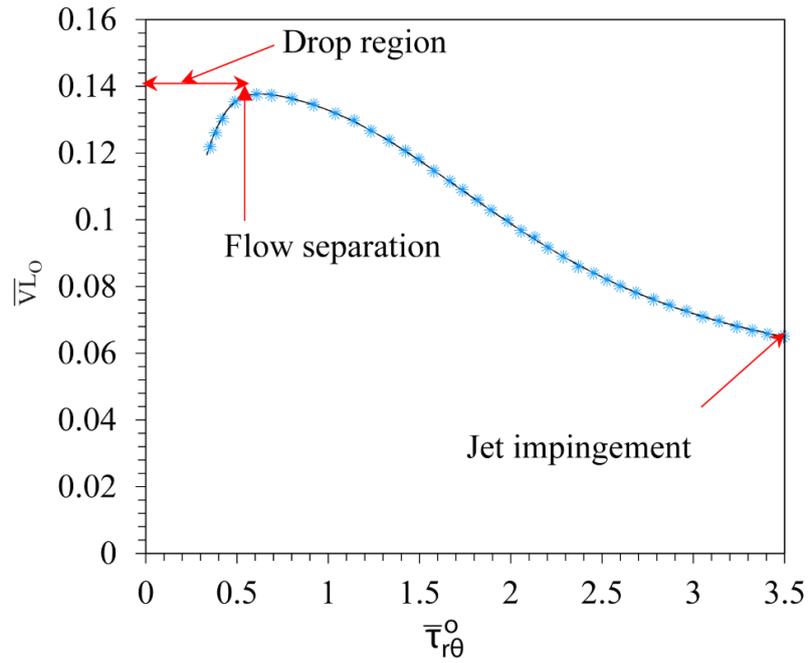

Figure 19: The relation between the normalized VL thickness ($\overline{VL}_o$) and normalized azimuthal wall shear stress ($\bar{\tau}_{r\theta}^o$) along axial direction at Re 375 for the outer wall.

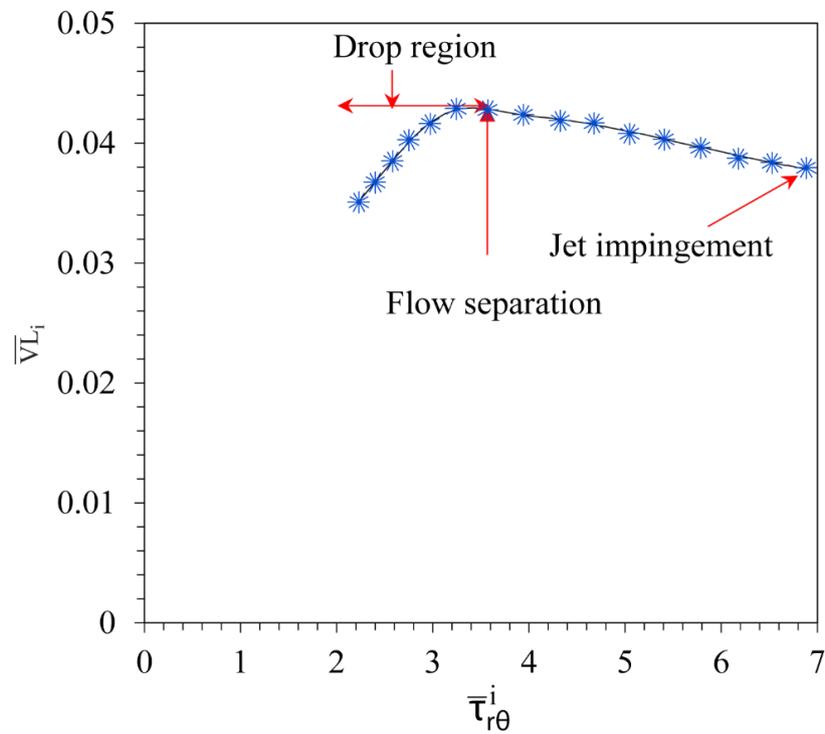

Figure 20: The relation between the normalized VL thickness ($\overline{VL}_i$) and azimuthal wall shear stress ($\bar{\tau}_{r\theta}^i$) along axial direction at Re 375 for the inner wall.

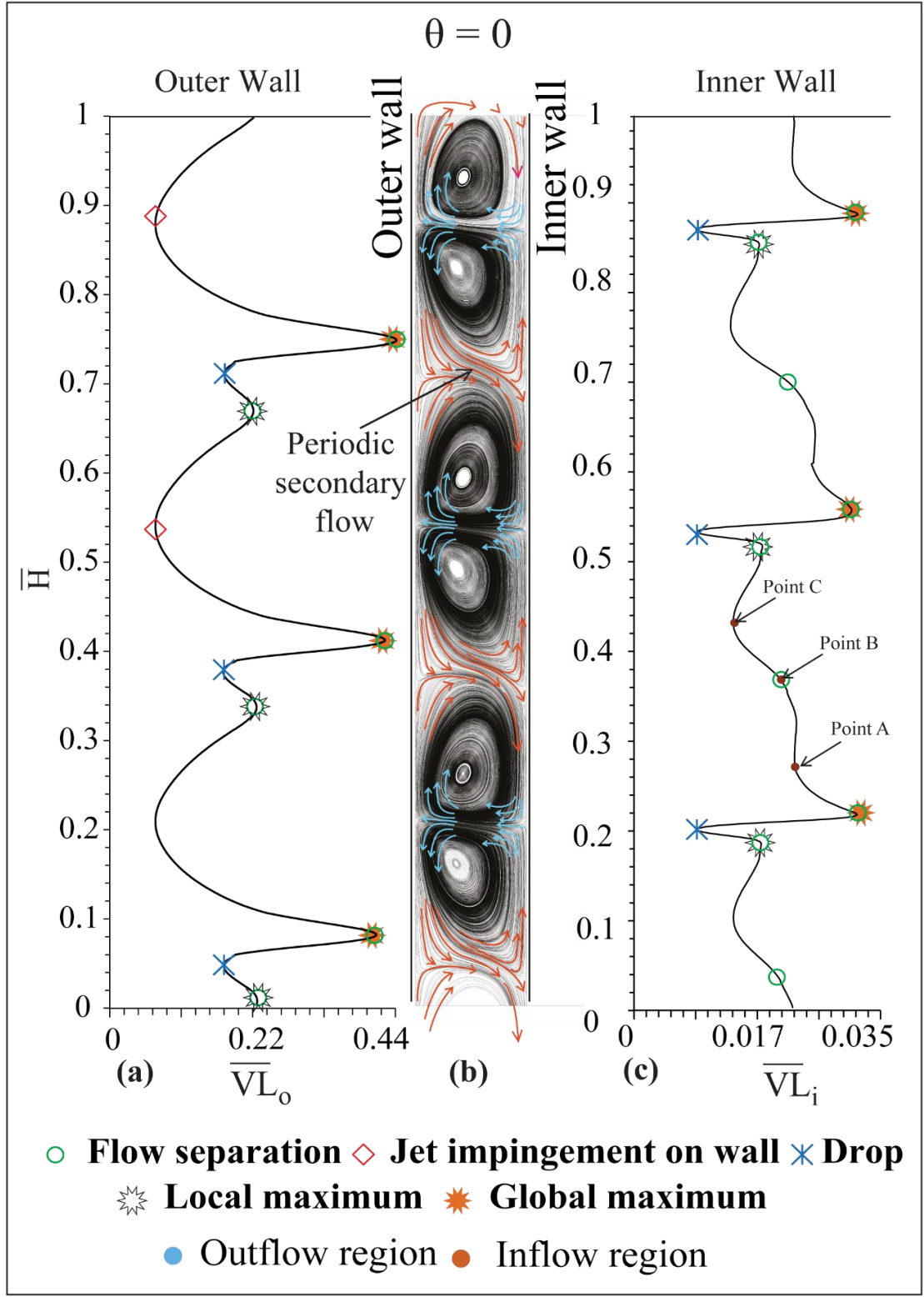

Figure 21: The distribution of normalized VL ($\overline{VL}$) in the flow structure observed during the transition phase of ATV into WVF along the axial direction for the inner and outer wall for Re 475 in r-Z plane at $\Theta=0$ (a) $\overline{VL}_o$ distribution at the outer wall (b) Streamlines associated with vortex structure (c) $\overline{VL}_i$ distribution at the inner wall.

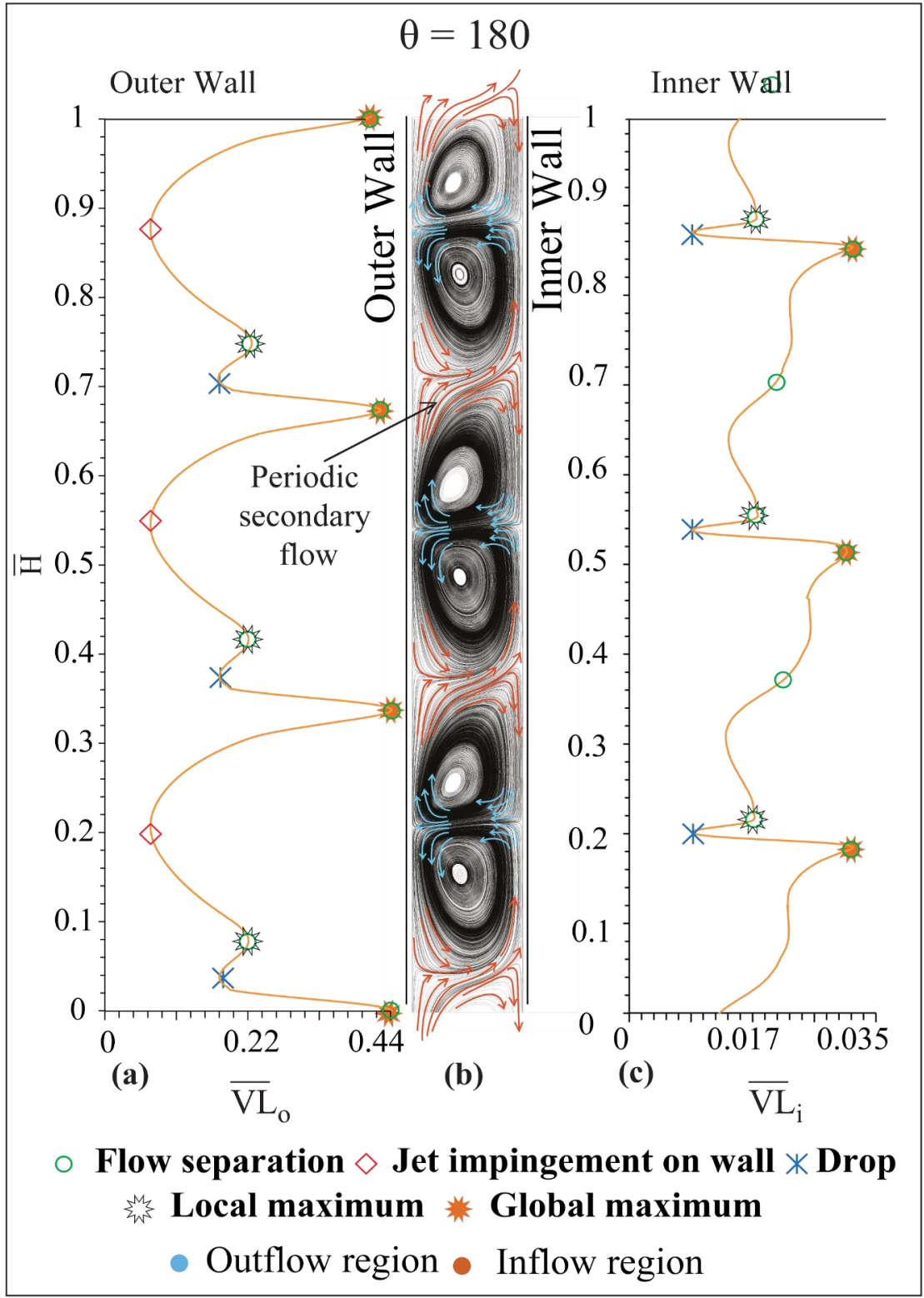

Figure 22: The distribution of normalized VL ($\overline{VL}$) in the flow structure observed during the transition phase of ATV into WVF along the axial direction for the inner and outer wall for Re 475 in r-Z plane at Θ= 180 (a) $\overline{VL}_o$ distribution at the outer wall (b) Streamlines associated with vortex structure (c) $\overline{VL}_i$ distribution at the inner wall.

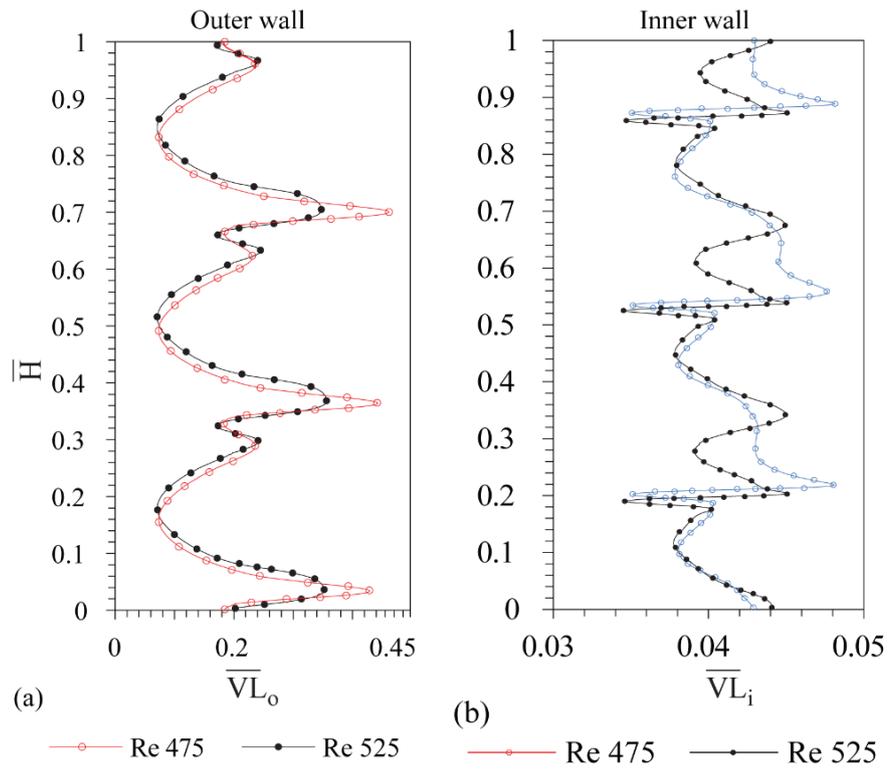

Figure 23: The spatial behaviour of normalized VL ($\overline{VL}$) along the axial direction in the transition of axisymmetric flow structure into non-axisymmetric flow structure with NNW 2.67 in the r-Z plane at Θ=0 (a) $\overline{VL_o}$ for the outer wall (b) $\overline{VL_i}$ for the inner wall.

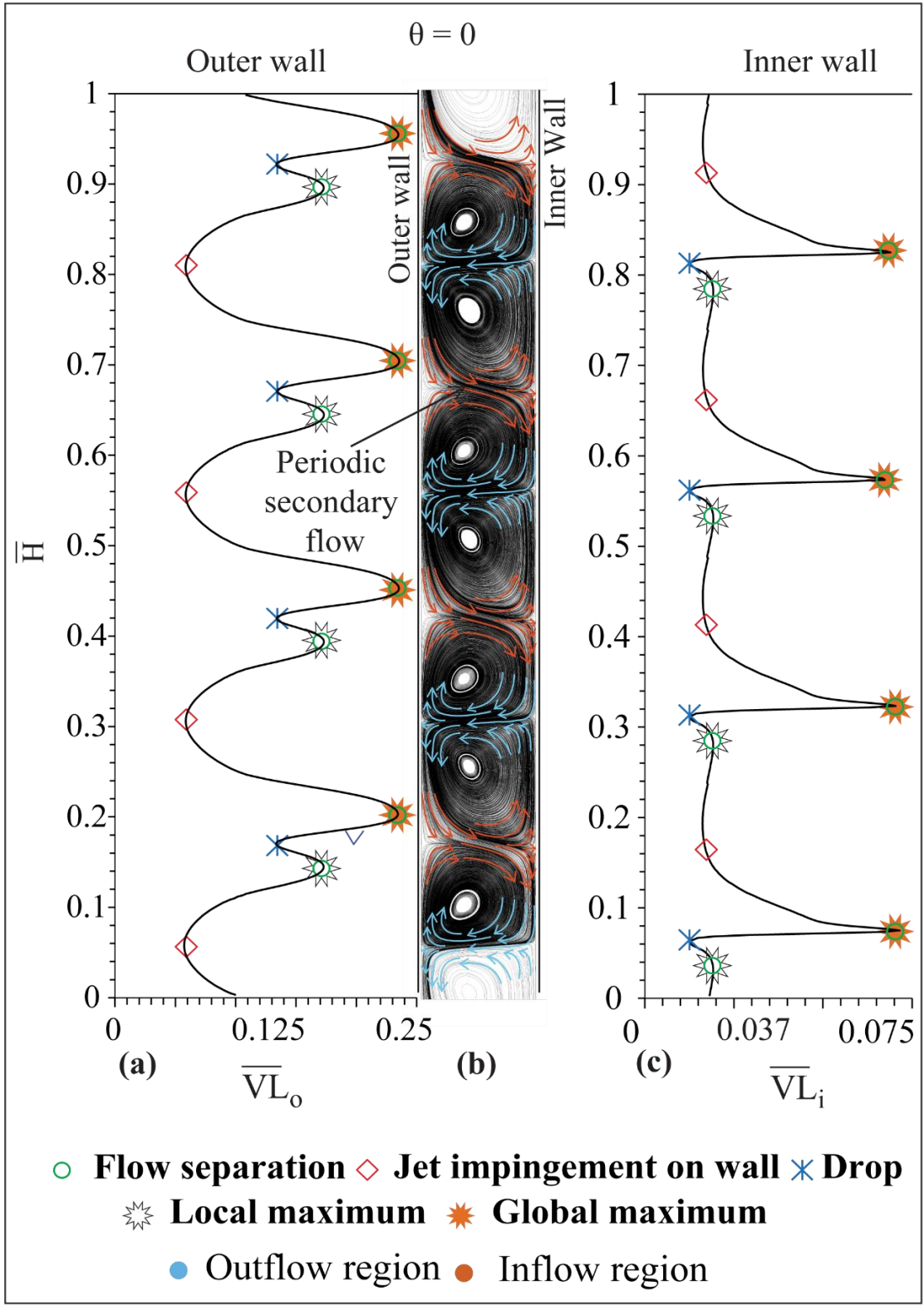

Figure 24: The distribution $\overline{VL}$ in the WVF along the axial direction for the inner and outer wall for Re 600 in the r-Z plane at $\Theta=0$ (a) $\overline{VL}_o$ distribution at the outer wall (b) Streamlines associated with vortex structure (c) $\overline{VL}_i$ distribution at the inner wall.

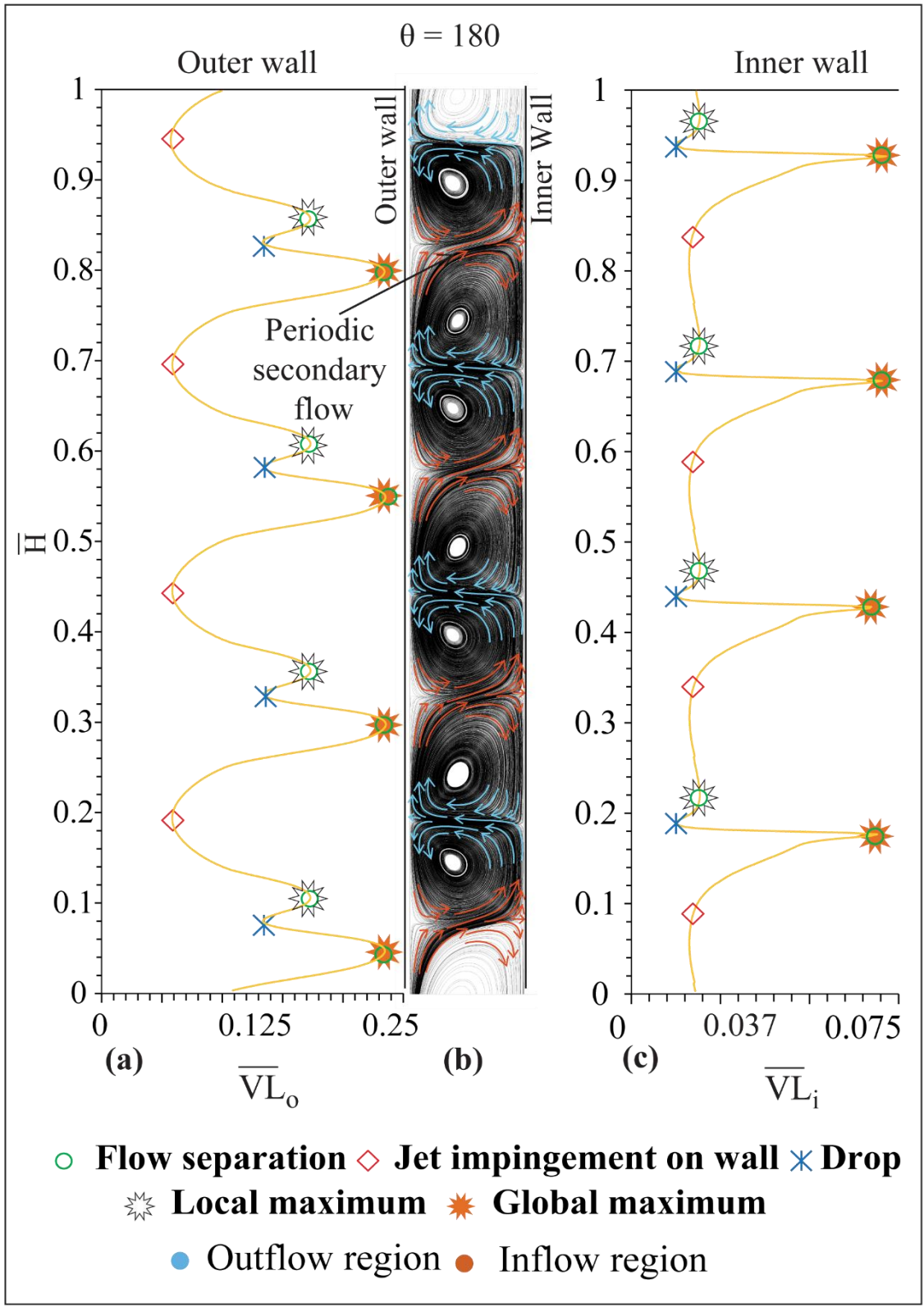

Figure 25: The distribution $\overline{VL}$ in the WVF along the axial direction for the inner and outer wall for Re 600 in the r-Z plane at $\Theta=180$ (a) $\overline{VL}_o$ distribution at the outer wall (b) Streamlines associated with vortex structure (c) $\overline{VL}_i$ distribution at the inner wall.

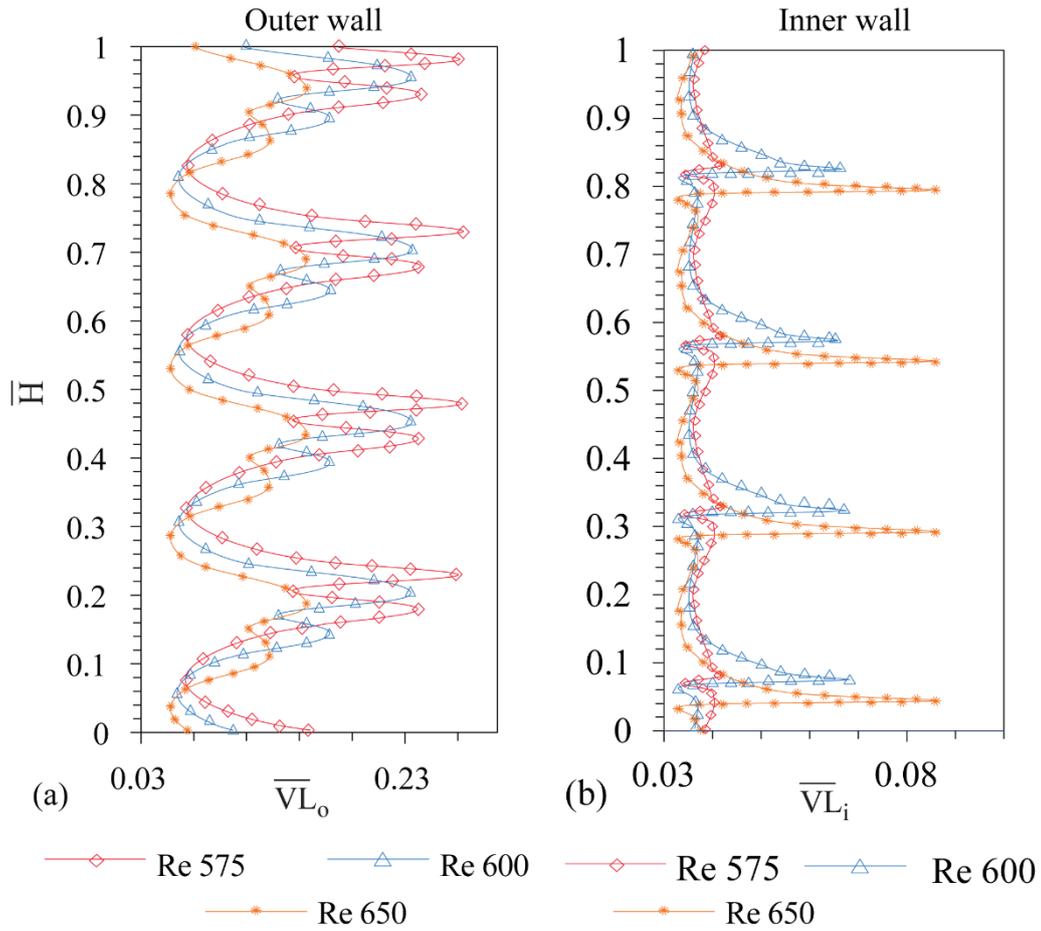

Figure 26: The spatial behaviour of normalized VL ($\overline{VL}$) along the axial direction in the WVF flow with NNW 2.00 in the r-Z plane at $\Theta=0$ (a) $\overline{VL}_o$ for the outer wall (b) $\overline{VL}_i$ for the inner wall.

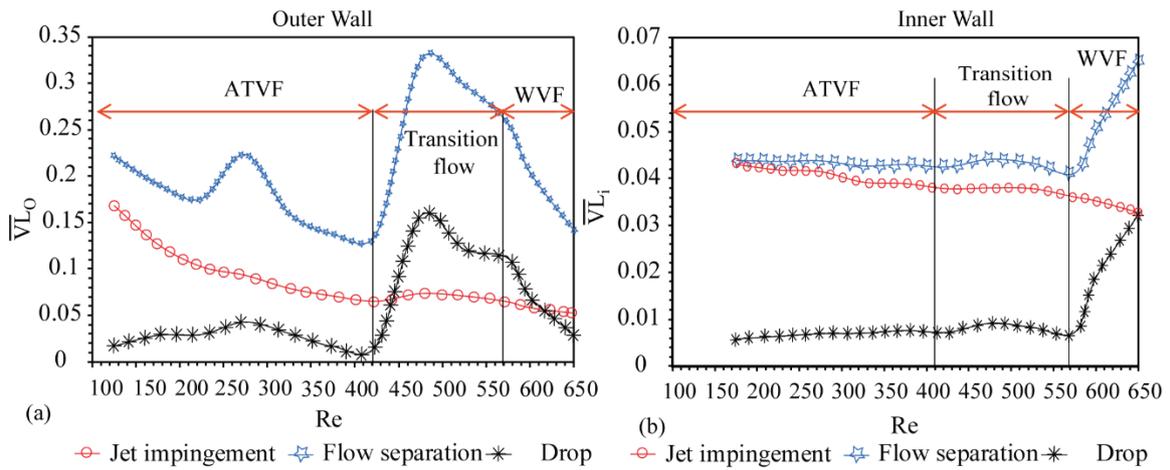

Figure 27: The behaviour of normalized VL with Re at jet impingement, flow separation and drop in the r-Z plane at $\Theta=0$ (a) the behaviour of VL thickness with Re at the outer wall (b) the behaviour of VL thickness with Re at the inner wall.

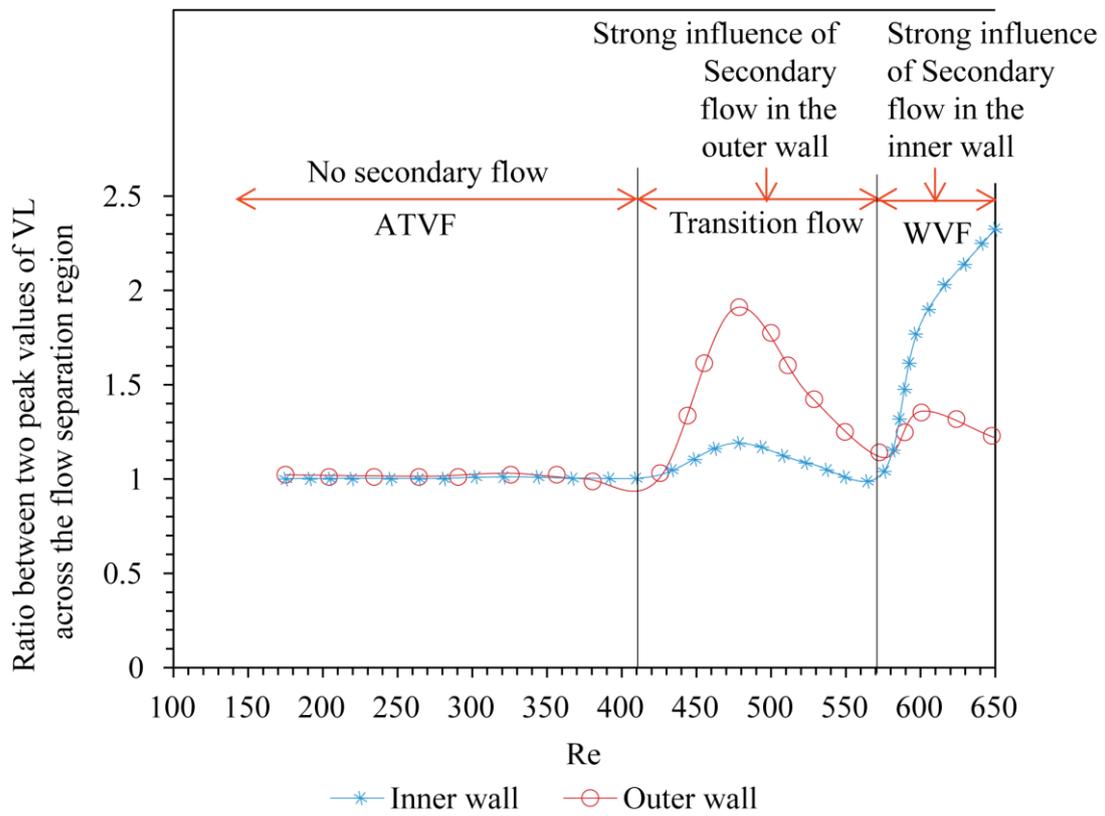

Figure 28: The variation of the ratio between two peak values of VL thickness across the flow separation region with Re in the r-Z plane at Ө=0.